\PassOptionsToPackage{unicode}{hyperref}
\PassOptionsToPackage{hyphens}{url}
\documentclass[
  11pt,
]{article}
\usepackage{lmodern}
\usepackage{amssymb,amsmath}
\usepackage{ifxetex,ifluatex}
\ifnum 0\ifxetex 1\fi\ifluatex 1\fi=0 
  \usepackage[T1]{fontenc}
  \usepackage[utf8]{inputenc}
  \usepackage{textcomp} 
\else 
  \usepackage{unicode-math}
  \defaultfontfeatures{Scale=MatchLowercase}
  \defaultfontfeatures[\rmfamily]{Ligatures=TeX,Scale=1}
\fi
\IfFileExists{upquote.sty}{\usepackage{upquote}}{}
\IfFileExists{microtype.sty}{
  \usepackage[]{microtype}
  \UseMicrotypeSet[protrusion]{basicmath} 
}{}
\makeatletter
\@ifundefined{KOMAClassName}{
  \IfFileExists{parskip.sty}{%
    \usepackage{parskip}
  }{
    \setlength{\parindent}{0pt}
    \setlength{\parskip}{6pt plus 2pt minus 1pt}}
}{
  \KOMAoptions{parskip=half}}
\makeatother
\usepackage{xcolor}
\IfFileExists{xurl.sty}{\usepackage{xurl}}{} 
\IfFileExists{bookmark.sty}{\usepackage{bookmark}}{\usepackage{hyperref}}
\hypersetup{
  pdftitle={Quantifying the Trendiness of Trends},
  hidelinks,
  pdfcreator={LaTeX via pandoc}}
\usepackage[margin=1in]{geometry}
\usepackage{graphicx,grffile}
\makeatletter
\def\maxwidth{\ifdim\Gin@nat@width>\linewidth\linewidth\else\Gin@nat@width\fi}
\def\maxheight{\ifdim\Gin@nat@height>\textheight\textheight\else\Gin@nat@height\fi}
\makeatother
\setkeys{Gin}{width=\maxwidth,height=\maxheight,keepaspectratio}
\makeatletter
\def\fps@figure{htbp}
\makeatother
\usepackage{bm}
\usepackage{amssymb}
\usepackage[labelfont=bf]{caption}

\usepackage{multirow}
\usepackage{float}
\floatstyle{plaintop}
\restylefloat{table}
\usepackage[amsthm,thmmarks]{ntheorem}

\newtheorem{assumption}{Assumption}
\newtheorem{proposition}{Proposition}
\theoremstyle{nonumberplain}

\usepackage{xcolor}

\usepackage{booktabs}
\usepackage{longtable}
\usepackage{array}
\usepackage{multirow}
\usepackage{wrapfig}
\usepackage{float}
\usepackage{colortbl}
\usepackage{pdflscape}
\usepackage{tabu}
\usepackage{threeparttable}
\usepackage{threeparttablex}
\usepackage[normalem]{ulem}
\usepackage{makecell}

\title{Quantifying the Trendiness of Trends}
\author{Andreas Kryger Jensen and Claus Thorn Ekstrøm\\
Biostatistics, Institute of Public Health, University of Copenhagen\\
\href{mailto:aeje@sund.ku.dk}{\nolinkurl{aeje@sund.ku.dk}},
\href{mailto:ekstrom@sund.ku.dk}{\nolinkurl{ekstrom@sund.ku.dk}}}
\date{03 October, 2020}

\begin{document}
\maketitle

\begin{abstract}
News media often report that the trend of some public health outcome has changed. These statements are frequently based on longitudinal data, and the change in trend is typically found to have occurred at the most recent data collection time point - if no change had occurred the story is less likely to be reported. Such claims may potentially influence public health decisions on a national level.

We propose two measures for quantifying the trendiness of trends. Assuming that reality evolves in continuous time we define what constitutes a trend and a change in trend, and introduce a probabilistic Trend Direction Index. This index has the interpretation of the probability that a latent characteristic has changed monotonicity at any given time conditional on observed data. We also define an index of Expected Trend Instability quantifying the expected number of changes in trend on an interval.

Using a latent Gaussian Process model we show how the Trend Direction Index and the Expected Trend Instability can be estimated in a Bayesian framework and use the methods to analyze the proportion of smokers in Denmark during the last 20 years, and the development of new COVID-19 cases in Italy from February 24th onwards.
\end{abstract}

\begin{center}
\end{center}

\hypertarget{introduction}{%
\section{Introduction}\label{introduction}}

Trend detection has received increased attention in many fields, and
while many important applications have their roots in the fields of
economics (stock development) and environmental change (global
temperature), trend identification has important ramifications in
industry (process monitoring), medicine (disease development) and public
health (changes in society).

\begin{figure}[htb]
\center\includegraphics{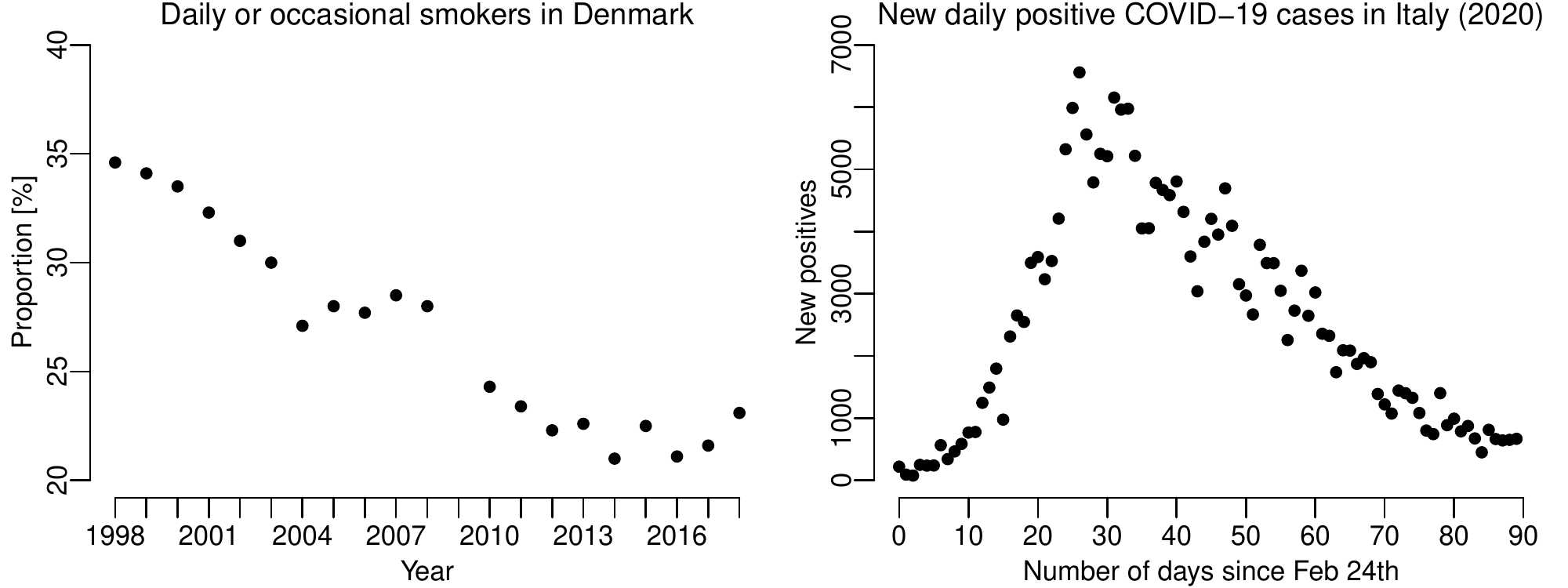}
\caption{Left panel: The proportion of daily or occasional smokers in Denmark during the last 20 years estimated from survey data and reported by the Danish Health Authority. The 2009 measurement is missing due to a problem with representativity. Right panel: The number of daily new cases tested positive for COVID-19 in Italy from February 24th, 2020 and 90 days onwards.}
\label{fig:rawDataPlot}
\end{figure}

This manuscript is concerned with the fundamental problem of estimating
an underlying trend based on random variables observed repeatedly over
time. In addition to this problem, we also wish to assess the
probabilities that such a trend is changing as a function of time. Our
motivation comes from two recent examples: in the first, the news media
in Denmark stated that the trend in the proportion of smokers in Denmark
had changed at the end of the year 2018 such that the proportion was now
increasing whereas it had been decreasing for the previous 20 years.
This statement was based on survey data collected yearly since 1998 and
reported by the Danish Health Authority (The Danish Health Authority
2019), and it is critical for the Danish Health Authorities to be able
to evaluate and react if an actual change in trend has occurred. The
second example relates to the recent outbreak of COVID-19 in Italy where
it is of tremendous importance to determine if the disease spread is
increasing or slowing down by considering the trend in number of new
cases (see Figure \ref{fig:rawDataPlot}).

The concept of ``trend'' is not itself well-defined and it is often up
to each researcher to define exactly what is meant by ``trend'' (Esterby
1993). Consequently, it is not obvious what constitutes a \emph{change}
in trend, and that makes it difficult to compare statistical methods to
detect trend changes, since they attempt to address slightly different
problems.

Change-point analysis is an often-used approach for detecting if a
change (of a prespecified type) has taken place (Basseville and
Nikiforov 1993; Barry and Hartigan 1993). However, change-point analysis
is marred by the fact that change-points that happen close to the
boundary of the observed time frame are notoriously difficult to detect,
and that they assume that change-points are abrupt changes that occur at
specific time point.

Another common approach to analyse trends in time series is to apply a
low-pass smoothing filter to the observed data in order to remove noise
and extract the underlying latent trend (Chandler and Scott 2011). But
in addition to the smoothing filter, it is also necessary to circumvent
the problem of what constitutes a change in trend so it is necessary to
specify a decision criterion to determine if any filtered changes are
relevant.

Gottlieb and Müller (2012) define a stickiness coefficient for
longitudinal data and use the stickiness coefficient to summarize the
extent to which deviations from the mean trajectory tend to co-vary over
time. While the stickiness coefficient determines changes from the
expected trend it is a measure that is not easily interpreted.

Kim et al. (2009) proposed an \(\ell_1\) Trend Filtering approach based
on an idea by Hodrick and Prescott (1997) that imposes sparsity on the
first order differences of the conditional mean. This produces trend
estimates that are piece-wise linear with the inherent assumption that
changes to the time series are abrupt. The Trend Filtering approach has
been further extended to sparsity of \(k\)th order differences and to a
Bayesian framework, where the flexibility of the difference of the
conditional mean is controlled through the prior distributions (Ramdas
and Tibshirani 2016; Kowal, Matteson, and Ruppert 2019)

We propose a new method for evaluating changes in the trend of a latent
function \(f\), which starts by a clear specification of the problem and
relevant measures of the trend: The Trend Direction Index gives a local
probability of the monotonicity of \(f\) and provides an answer to the
research question ``What is the probability that the latent function is
increasing?'' The Expected Trend Instability index gives the expected
number of changes in the monotonicity of \(f\) over a time interval, and
thus provides information about the volatility of \(f\). It can be used
to evaluate research questions about the rate of change in direction of
\(f\). In particular, it can be used to gauge how surprising such
statements as "It is the first time in 20 years that \(f\) has stopped
decreasing and started to increase'' is. Besides providing exact answers
to specific and relevant research questions, our proposed method
estimates the actual probability that the trend is changing.

The manuscript is structured as follows: In Section \ref{sec:method} we
present our statistical model based on a latent Gaussian Process
formulation giving rise to explicit expressions for the Trend Direction
Index and the Expected Trend Instability conditional on observed data.
Section \ref{sec:estimation} is concerned with estimating the models
parameters. In Section \ref{sec:simulation} we undertake a simulation
study to show the performance of the proposed method and to compare it
with Trend Filtering, and in Section \ref{sec:application} we provide
extended applications to our two cases: the development of the
proportion of smokers in Denmark during the last 20 years, and the
development of new COVID-19 cases in Italy. We conclude with a
discussion.

Proofs of propositions are given in the Supplementary Material, and
reproducible code and Stan implementations of the model are available at
the first author's GitHub repository (Jensen 2019).

\hypertarget{sec:method}{%
\section{Methods}\label{sec:method}}

We assume that reality evolves in continuous time
\(t \in \mathcal{T} \subset \mathbb{R}\) and that there exists a random,
latent function \(f = \left\{f(t) : t \in \mathcal{T}\right\}\) with
hyper-parameters \(\bm{\Theta}\) sufficiently smooth on a compact subset
of the real line, \(\mathcal{T}\), governing the underlying evolution of
some observable characteristic in a population. We can observe this
latent characteristic with noise by sampling \(f\) at discrete time
points according to the additive model \(Y_i = f(t_i) + \varepsilon_i\)
where \(\varepsilon_i\) is a zero mean random variable independent of
\(f(t_i)\). Given observations of the form \((Y_i, t_i)_{i=1}^n\) we are
interested in modeling the dynamical properties of \(f\).

The trend of \(f\) is defined as its instantaneous slope given by the
function \(df(t) = \left(\frac{\mathrm{d}f(s)}{\mathrm{d}s}\right)(t)\),
and \(f\) is increasing and has a positive trend at \(t\) if
\(df(t) > 0\), and \(f\) is decreasing with a negative trend at \(t\) if
\(df(t) < 0\). A change in trend is defined as a change in the sign of
\(df\), i.e., when \(f\) goes from increasing to decreasing or vice
versa.

As \(f\) is a random function inferred by observations in discrete time
there are no singular points where a sign change in the trend can be
asserted almost surely from the probability distribution of the estimate
of \(f\). This is instead characterized by a gradual and continuous
change in the probability of the monotonicity of \(f\), and an
assessment of a change in trend is defined by the probability of the
sign of \(df\). This stands in contrast to traditional change-point
models which assume that there are one or more exact time points where a
sudden change in a function or its parameterization occurs (Carlstein,
Müller, and Siegmund 1994).

The probability of a positive trend for \(f\) at time \(t + \delta\) is
quantified by the Trend Direction Index \begin{align}
  \mathrm{TDI}(t, \delta) = P(df(t + \delta ) > 0 \mid \mathcal{F}_t), \quad t \in \mathcal{T}\label{eq:TCIdef}
\end{align} where \(\mathcal{F}_t\) is a \(\sigma\)-algebra of available
information observed up until time \(t\). The value of
\(\mathrm{TDI}(t, \delta)\) is a local probabilistic index, and it is
equal to the probability that \(f\) is increasing at time \(t + \delta\)
given everything known about the data generating process up until and
including time \(t\). A similar definition can be given for a negative
trend but that is equal to \(1 - \mathrm{TDI}(t, \delta)\) and therefore
redundant. The sign of \(\delta\) determines whether the Trend Direction
Index estimates the past (\(\delta \leq 0\)) or forecasts the future
(\(\delta > 0\)). Most of the examples seen in the news concerning
public health outcomes are concerned with \(t\) being equal to the
current calendar time and \(\delta = 0\). This excludes the usage of
both change-point and segmented regression models (Quandt 1958) as there
are no observations available beyond the stipulated change-point. A
useful reparameterization of the Trend Direction Index is
\(\mathrm{TDI}(\max \mathcal{T}, t - \max \mathcal{T})\) with
\(t \leq \max \mathcal{T}\). This parameterization conditions on the
full observation period and looks back in time whereas setting
\(t = \max \mathcal{T}\) corresponds to the current Trend Direction
Index at the end of the observation period.

In addition to the Trend Direction Index we define a global measure of
trend instability. Informally, we say that a random function \(f\) is
\textit{trend stable} on an interval \(\mathcal{I}\) if its sample paths
maintain their monotonicity so that the trends do not change sign on the
interval. To quantify the trend \emph{in}stability we propose to use the
expected number of zero-crossings by \(df\) on \(\mathcal{I}\). We
define the Expected Trend Instability as \begin{align}
  \text{ETI}(\mathcal{I}) = \mathop{\mathrm{E}}\left[\#\left\{t \in \mathcal{I} : df(t) = 0\right\} \mid \mathcal{F}\right]\label{eq:ETIdef}
\end{align} equal to the expected value of the size of the random set of
zero-crossings by \(df\) on \(\mathcal{I}\) when conditioning on a
suitable \(\sigma\)-algebra \(\mathcal{F}\). A common case is when
\(\mathcal{F}\) is generated by all observed data on \(\mathcal{T}\) and
\(\mathcal{I} \subseteq \mathcal{T}\). The lower
\(\text{ETI}(\mathcal{I})\) is, the more stable the trend of \(f\) on
\(\mathcal{I}\) is and vice versa.

We note, thanks to a comment by an anonymous reviewer, that the Expected
Trend Instability represents an upper bound for the probability of
observing at least one zero-crossing event, \(df(\cdot) = 0\), on
\(\mathcal{I}\) since \begin{align*}
  \text{ETI}(\mathcal{I}) &= \sum_{k=0}^\infty P(\# \{t \in \mathcal{I} : df(t) = 0\} = k) \cdot k\\
  &\geq \sum_{k=1}^\infty P(\# \{t \in \mathcal{I} : df(t) = 0\} = k)\\
  &= P(\# \{t \in \mathcal{I} : df(t) \geq 0\} \geq 1)
\end{align*} where the inequality becomes sharp if
\(\sum_{k\geq2} P(\# \{t \in \mathcal{I} : df(t) = 0\} = k)\) is small.
The Expected Trend Instability can therefore be used over smaller
intervals \(\mathcal{I}\) to provide statements about the expected
probability that a change in trend will happen.

These two general definitions of trendiness will be evaluated in the
light of a particular statistical model in the next sections leading to
expressions of their estimates.

\hypertarget{latent-gaussian-process-model}{%
\subsection{Latent Gaussian Process
Model}\label{latent-gaussian-process-model}}

The definitions given in the previous section impose restrictions on the
latent function \(f\). We shall assume that \(f\) is a Gaussian Process
on \(\mathcal{T}\). From a Bayesian perspective this is equivalent to
imposing an infinite dimensional prior distribution on the latent
characteristic governing the observed outcomes. Statistical models with
Gaussian Process priors are a flexible approach for non-parametric
regression (Radford 1999), and using a latent Gaussian Process provides
an analytically tractable way for performing statistical inference on
its derivatives. The general idea of our model is to apply the
properties of the Gaussian Process prior on the latent characteristic to
update the finite dimensional distributions by conditioning on the
observed data. This results in a joint posterior Gaussian Process for
\(f\) and its derivatives from which estimates of the trend indices can
be derived.

A random function \(f\) is a Gaussian Process if and only if the vector
\((f(t_1), \ldots, f(t_n))\) has a multivariate normal distribution for
every finite set of evaluation points \((t_1, \ldots, t_n)\), and we
write \(f \sim \mathcal{GP}(\mu(\cdot), C(\cdot, \cdot))\) where \(\mu\)
is the mean function, and \(C\) is a symmetric, positive definite
covariance function (Cramer and Leadbetter 1967). We observed dependent
data in terms of outcomes and their associated sampling times,
\((Y_i, t_i)_{i=1}^n\), and we assume that the data are generated by the
hierarchical model \begin{align}
\begin{split}
  f \mid \bm{\beta}, \bm{\theta} &\sim \mathcal{GP}(\mu_{\bm{\beta}}(\cdot), C_{\bm{\theta}}(\cdot,\cdot))\\
  Y_i \mid t_i, f(t_i), \bm{\Theta} &\overset{iid}{\sim} N(f(t_i), \sigma^2), \quad \bm{\Theta} = (\bm{\beta}, \bm{\theta}, \sigma)
\end{split}
\label{eq:generatingProcess}
\end{align} where \(\bm{\beta}\) is a vector of parameters for the mean
function of \(f\), \(\bm{\theta}\) is a vector of parameters governing
the covariance of \(f\), and \(\sigma\) is the conditional standard
deviation of the observations. Together, these parameters,
\(\bm{\Theta}\), are hyper-parameters of the model.

\vspace{0.2cm}

\begin{assumption}
We assume the following regularity conditions.
\begin{enumerate}
  \item[A1:]{$f$ is a separable Gaussian Process.}
  \item[A2:]{$\mathop{\mathrm{E}}[f(t) \mid \bm{\beta}, \bm{\theta}] = \mu_{\bm{\beta}}(t)$ is a twice continuously differentiable function.}
  \item[A3:]{$\mathop{\mathrm{Cov}}[f(s), f(t) \mid \bm{\beta}, \bm{\theta}] = C_{\bm{\theta}}(s,t)$ has mixed third-order partial derivatives continuous at the diagonal.}
  \item[A4:]{The joint distribution of $(df(t), d^2\!f(t) \mid \bm{\beta}, \bm{\theta})$ is non-degenerate at any $t$ i.e., $\mathop{\mathrm{Cov}}[df(t), df(t) \mid \bm{\beta}, \bm{\theta}] > 0$ and $-1 < \mathop{\mathrm{Cor}}[df(t), d^2\!f(t) \mid \bm{\beta}, \bm{\theta}] < 1$.}
\end{enumerate}
\label{assumptions}
\end{assumption}

Assumption A1 is a technical condition required to ensure that
functionals of \(f\) defined on an uncountable index set such as
\(\mathcal{T}\) can form random variables. All continuous Gaussian
Processes are separable. A2 is required in order to make
\(\mathop{\mathrm{E}}[df \mid \bm{\beta}, \bm{\theta}]\) and
\(\mathop{\mathrm{E}}[d^2\!f \mid \bm{\beta}, \bm{\theta}]\)
well-defined as expected values for the joint distribution in Equation
(\ref{eq:latentJoint}). This is a modeling choice and can be fulfilled
by choosing \(\mu_{\bm{\beta}}\) accordingly. A3 is similarly required
in order to make the prior covariance matrix in Equation
(\ref{eq:latentJoint}) well-defined. We discuss practicalities regarding
this assumption in Section \ref{sec:priorparametrization}. A4 states
that one cannot use a covariance function for \(f\) where i) the
covariance of \(df\) becomes degenerate with no variability or ii) a
covariance function for which the first and second order derivatives are
perfectly correlated. This is to ensure that the Trend Direction Index
and the Expected Trend Instability are well-defined quantities. This is
again a modeling choice and can always be verified in practice for a
particular choice of \(C_{\bm{\theta}}\). It should be noted that while
assumptions A2 and A3 impose restrictions on the functions that can be
used to describe the mean and the covariance functions, the class of
functions is still extremely flexible and should accommodate most
modeling situations.

Under the above assumptions an important property of a Gaussian Process
is that it together with its first and second derivatives is a
multivariate Gaussian Process with explicit expressions for the joint
mean, covariance and cross-covariance functions. Specifically, the joint
distribution of the latent function, \(f\), and its first and second
derivatives, \(df\) and \(d^2\!f\), is the multivariate Gaussian Process
\begin{align}
  \begin{bmatrix}f(s)\\ df(t)\\ d^2\!f(u)\end{bmatrix} \mid \bm{\beta}, \bm{\theta} &\sim \mathcal{GP}\left(\begin{bmatrix}\mu_{\bm{\beta}}(s)\\ d\mu_{\bm{\beta}}(t)\\ d^2\!\mu_{\bm{\beta}}(u)\end{bmatrix}, \begin{bmatrix}C_{\bm{\theta}}(s, s^\prime) & \partial_2 C_{\bm{\theta}}(s, t) & \partial_2^2 C_{\bm{\theta}}(s, u)\\ \partial_1 C_{\bm{\theta}}(t, s) & \partial_1 \partial_2 C_{\bm{\theta}}(t, t^\prime) & \partial_1 \partial_2^2 C_{\bm{\theta}}(t, u)\\ \partial_1^2 C_{\bm{\theta}}(u, s) & \partial_1^2\partial_2 C_{\bm{\theta}}(u, t) & \partial_1^2 \partial_2^2 C_{\bm{\theta}}(u, u^\prime)\end{bmatrix}\right)\label{eq:latentJoint}
\end{align} where \(d^k\!\mu_{\bm{\beta}}\) is the \(k\)'th derivative
of \(\mu_{\bm{\beta}}\) and \(\partial_j^k\) denotes the \(k\)'th order
partial derivative with respect to the \(j\)'th variable (Cramer and
Leadbetter 1967). Proposition \ref{prop:GPposterior} states the joint
posterior distribution of \((f, df, d^2\!f)\) conditional on the
observed data.

\vspace{0.2cm}

\begin{proposition}
Let the data generating model be defined as in Equation (\ref{eq:generatingProcess}) and $\mathbf{Y} = (Y_1, \ldots, Y_n)$ the vector of observed outcomes together with its sampling times $\mathbf{t} = (t_1, \ldots, t_n)$. Then by the conditions in Assumption \ref{assumptions} the joint distribution of $(f, df, d^2\!f)$ conditional on $\mathbf{Y}$, $\mathbf{t}$ and the hyper-parameters $\bm{\Theta}$ evaluated at any finite vector $\mathbf{t}^\ast$ of $p$ time points is
\begin{align*}
\begin{bmatrix}f(\mathbf{t}^\ast)\\ df(\mathbf{t}^\ast)\\ d^2\!f(\mathbf{t}^\ast)\end{bmatrix} \mid \mathbf{Y}, \mathbf{t}, \bm{\Theta} \sim N\left(\bm{\mu},  \bm{\Sigma}\right)
\end{align*}
where $\bm{\mu} \in \mathbb{R}^{3p}$ is the column vector of posterior expectations and $\bm{\Sigma} \in \mathbb{R}^{3p \times 3p}$ is the joint posterior covariance matrix. Partitioning these as
\begin{align*}
  \bm{\mu} = \begin{bmatrix}\mu_{f}(\mathbf{t^\ast} \mid \bm{\Theta})\\ \mu_{df}(\mathbf{t^\ast} \mid \bm{\Theta})\\ \mu_{d^2\!f}(\mathbf{t^\ast} \mid \bm{\Theta})\end{bmatrix}, \quad \bm{\Sigma} = \begin{bmatrix}\Sigma_{f}(\mathbf{t^\ast},\mathbf{t^\ast} \mid \bm{\Theta}) &  \Sigma_{f,df}(\mathbf{t^\ast},\mathbf{t^\ast} \mid \bm{\Theta}) & \Sigma_{f,d^2\!f}(\mathbf{t^\ast},\mathbf{t^\ast} \mid \bm{\Theta})\\  \Sigma_{f,df}(\mathbf{t^\ast},\mathbf{t^\ast} \mid \bm{\Theta})^T & \Sigma_{df}(\mathbf{t^\ast},\mathbf{t^\ast} \mid \bm{\Theta}) & \Sigma_{df,d^2\!f}(\mathbf{t^\ast},\mathbf{t^\ast} \mid \bm{\Theta})\\ \Sigma_{f,d^2\!f}(\mathbf{t^\ast}, \mathbf{t^\ast} \mid \bm{\Theta})^T & \Sigma_{df,d^2\!f}(\mathbf{t^\ast}, \mathbf{t^\ast} \mid \bm{\Theta})^T & \Sigma_{d^2\!f}(\mathbf{t^\ast}, \mathbf{t^\ast} \mid \bm{\Theta})\end{bmatrix} 
\end{align*}
the individual components are given by
\begin{align*}
  \mu_{f}(\mathbf{t}^\ast \mid \bm{\Theta}) &= \mu_{\bm{\beta}}(\mathbf{t}^\ast) + C_{\bm{\theta}}(\mathbf{t}^\ast, \mathbf{t})\left(C_{\bm{\theta}}(\mathbf{t}, \mathbf{t}) + \sigma^2 I\right)^{-1}\left(\mathbf{Y} - \mu_{\bm{\beta}}(\mathbf{t})\right)\\
  \mu_{df}(\mathbf{t}^\ast \mid \bm{\Theta}) &= d\mu_{\bm{\beta}}(\mathbf{t}^\ast) + \partial_1 C_{\bm{\theta}}(\mathbf{t}^\ast, \mathbf{t})\left(C_{\bm{\theta}}(\mathbf{t}, \mathbf{t}) + \sigma^2 I\right)^{-1}\left(\mathbf{Y} - \mu_{\bm{\beta}}(\mathbf{t})\right) \\
  \mu_{d^2\!f}(\mathbf{t}^\ast \mid \bm{\Theta}) &= d^2\!\mu_{\bm{\beta}}(\mathbf{t}^\ast) + \partial_1^2 C_{\bm{\theta}}(\mathbf{t}^\ast, \mathbf{t})\left(C_{\bm{\theta}}(\mathbf{t}, \mathbf{t}) + \sigma^2 I\right)^{-1}\left(\mathbf{Y} - \mu_{\bm{\beta}}(\mathbf{t})\right)\\
  \Sigma_{f}(\mathbf{t}^\ast, \mathbf{t}^\ast \mid \bm{\Theta}) &= C_{\bm{\theta}}(\mathbf{t}^\ast, \mathbf{t}^\ast) - C_{\bm{\theta}}(\mathbf{t}^\ast, \mathbf{t})\left(C_{\bm{\theta}}(\mathbf{t}, \mathbf{t}) + \sigma^2 I\right)^{-1} C_{\bm{\theta}}(\mathbf{t}, \mathbf{t}^\ast)\\
  \Sigma_{df}(\mathbf{t}^\ast, \mathbf{t}^\ast \mid \bm{\Theta}) &= \partial_1\partial_2C_{\bm{\theta}}(\mathbf{t}^\ast, \mathbf{t}^\ast) - \partial_1C_{\bm{\theta}}(\mathbf{t}^\ast, \mathbf{t})\left(C_{\bm{\theta}}(\mathbf{t}, \mathbf{t}) + \sigma^2 I\right)^{-1} \partial_2C_{\bm{\theta}}(\mathbf{t}, \mathbf{t}^\ast)\\
  \Sigma_{d^2\!f}(\mathbf{t}^\ast, \mathbf{t}^\ast \mid \bm{\Theta}) &= \partial_1^2\partial_2^2 C_{\bm{\theta}}(\mathbf{t}^\ast, \mathbf{t}^\ast) - \partial_1^2 C_{\bm{\theta}}(\mathbf{t}^\ast, \mathbf{t})\left(C_{\bm{\theta}}(\mathbf{t}, \mathbf{t}) + \sigma^2 I\right)^{-1} \partial_2^2 C_{\bm{\theta}}(\mathbf{t}, \mathbf{t}^\ast)\\
  \Sigma_{f, df}(\mathbf{t}^\ast, \mathbf{t}^\ast \mid \bm{\Theta}) &= \partial_2 C_{\bm{\theta}}(\mathbf{t}^\ast, \mathbf{t}^\ast) - C_{\bm{\theta}}(\mathbf{t}^\ast, \mathbf{t})\left(C_{\bm{\theta}}(\mathbf{t}, \mathbf{t}) + \sigma^2 I\right)^{-1} \partial_2 C_{\bm{\theta}}(\mathbf{t}, \mathbf{t}^\ast)\\
  \Sigma_{f, d^2\!f}(\mathbf{t}^\ast, \mathbf{t}^\ast \mid \bm{\Theta}) &= \partial_2^2 C_{\bm{\theta}}(\mathbf{t}^\ast, \mathbf{t}^\ast) - C_{\bm{\theta}}(\mathbf{t}^\ast, \mathbf{t})\left(C_{\bm{\theta}}(\mathbf{t}, \mathbf{t}) + \sigma^2 I\right)^{-1} \partial_2^2 C_{\bm{\theta}}(\mathbf{t}, \mathbf{t}^\ast)\\
  \Sigma_{df, d^2\!f}(\mathbf{t}^\ast, \mathbf{t}^\ast \mid \bm{\Theta}) &= \partial_1 \partial_2^2 C_{\bm{\theta}}(\mathbf{t}^\ast, \mathbf{t}^\ast) - \partial_1 C_{\bm{\theta}}(\mathbf{t}^\ast, \mathbf{t})\left(C_{\bm{\theta}}(\mathbf{t}, \mathbf{t}) + \sigma^2 I\right)^{-1} \partial_2^2 C_{\bm{\theta}}(\mathbf{t}, \mathbf{t}^\ast)
\end{align*}
\label{prop:GPposterior}
\end{proposition}

In addition to the closed-form expressions of the posterior
distributions of the latent function and its derivatives given in
Proposition \ref{prop:GPposterior} it can also be useful to consider the
predictive distribution of the model corresponding to the conditional
distribution of a new observations given the observed data. This is
given by \begin{align}
Y^\ast(t^\ast) \mid t^\ast, \mathbf{Y}, \mathbf{t}, \bm{\Theta} \sim N\left(\mu_{f}(t^\ast \mid \bm{\Theta}), \Sigma_{f}(t^\ast, t^\ast \mid \bm{\Theta}) + \sigma^2\right)\label{eq:PPD}
\end{align} where \(Y^\ast\) is a new random variable predicted at some
arbitrary time point \(t^\ast\).

Proposition \ref{prop:GPposterior} grants the foundation for the rest of
the methodological development. While the Gaussian Process prior might
seem restrictive it has the computational advantage that the posterior
of \((f, df, d^2\!f)\) is characterized by the finite dimensional joint
distributions which in turn are given by the mean and covariance
functions. Further, for fixed \(\Theta\) there exists one-step update
formulas for the posteriors that can be useful and efficient for
real-time or online applications. A theoretical result motivating the
applicability is that the model possesses the universal approximation
property meaning that it can approximate any continuous function
uniformly on a closed interval of the real line to any desired tolerance
given sufficient data (Micchelli, Xu, and Zhang 2006).

In the following two subsections we show how the Trend Direction Index
and the Expected Trend Instability can be expressed under the data
generating model in Equation (\ref{eq:generatingProcess}) using the
results of Proposition \ref{prop:GPposterior}.

\hypertarget{the-trend-direction-index}{%
\subsection{The Trend Direction Index}\label{the-trend-direction-index}}

The Trend Direction Index was defined generally in Equation
(\ref{eq:TCIdef}). Conditioning on the \(\sigma\)-algebra of all
observed data we may express the Trend Direction Index under the model
in Equation (\ref{eq:generatingProcess}) through the posterior
distribution of \(df\). The following proposition states this result.

\vspace{0.2cm}

\begin{proposition}
Let $\mathcal{F}_t$ be the $\sigma$-algebra generated by $(\mathbf{Y}, \mathbf{t})$ where $\mathbf{Y} = (Y_1, \ldots, Y_n)$ is the vector of observed outcomes and $\mathbf{t} = (t_1, \ldots, t_n)$ the associated sampling times. Furthermore, assume that assumptions A1-A3 above are fulfilled. The Trend Direction Index defined in Equation (\ref{eq:TCIdef}) can then be written in terms of the posterior distribution of $df$ as
\begin{align*}
  \mathrm{TDI}(t, \delta \mid \bm{\Theta}) &= P(df(t + \delta ) > 0 \mid \mathbf{Y}, \mathbf{t}, \bm{\Theta})\\
  &= \frac{1}{2} + \frac{1}{2}\mathop{\mathrm{Erf}}\left(\frac{\mu_{df}(t + \delta \mid \bm{\Theta})}{2^{1/2}\Sigma_{df}(t + \delta, t + \delta \mid \bm{\Theta})^{1/2}}\right)
\end{align*}
where $\mathop{\mathrm{Erf}}\colon\, x \mapsto 2\pi^{-1/2}\int_0^x \exp(-u^2)\mathrm{d}u$ is the error function and $\mu_{df}$ and $\Sigma_{df}$ are the posterior mean and covariance of the trend defined in Proposition \ref{prop:GPposterior}.
\label{prop:TDIposterior}
\end{proposition}

Proposition \ref{prop:TDIposterior} shows that the Trend Direction Index
is equal to \(0.5\) when \(\mu_{df}(t + \delta \mid \bm{\Theta}) = 0\)
corresponding to when the expected value of the posterior of \(f\) is
locally constant at time \(t + \delta\). A decision rule based on
\(\mathrm{TDI}(t, \delta \mid \bm{\Theta}) \lessgtr 50\%\) is therefore
a natural choice for assessing the local trendiness of \(f\). However,
different thresholds based on external loss or utility functions can be
used depending on the application. Note, that the requirements of
assumptions A2 and A3 can be reduced to first order differentiable (A2)
and mixed second order differentiability (A3) respectively if only the
TDI is to be used.

\begin{figure}[htb]
\center\includegraphics{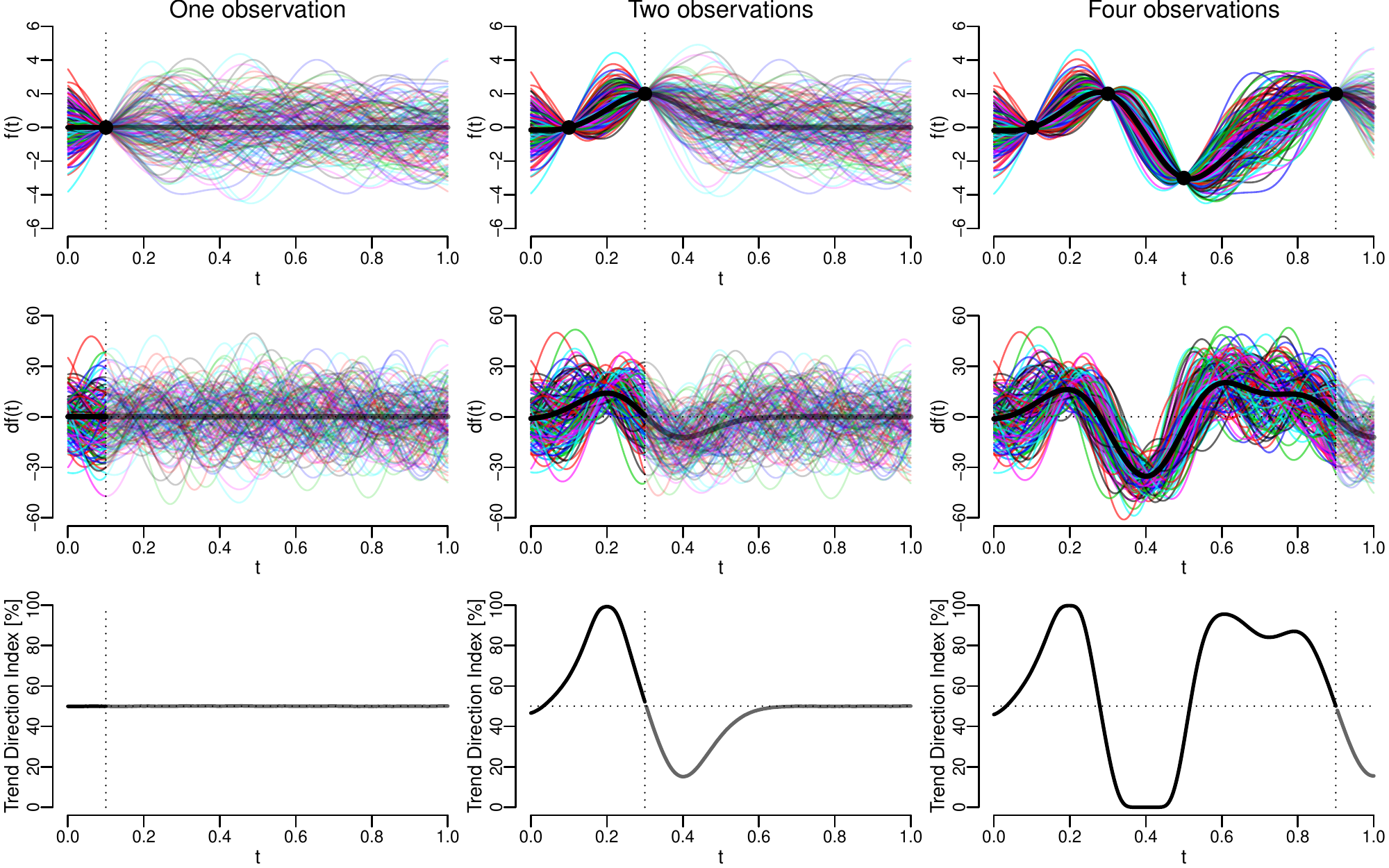}
\caption{150 realizations from the posterior distribution of $f$ (top row), $df$ (middle row) with expected values in bold and the Trend Direction Index (bottom row) conditional on one, two and four noise free observations. Dotted vertical lines show the points in time after which forecasting takes place.}
\label{fig:probabilisticExample}
\end{figure}

The Trend Direction Index is illustrated in Figure
\ref{fig:probabilisticExample} in the noise free case, \(\sigma = 0\),
with known values of \(\bm{\beta}\) and \(\bm{\theta}\), and where the
prior expected value of \(f\) is set equal to
\(\mu_{\bm{\beta}}(t) = 0\). The first and second rows of the plot show
150 random sample paths from the posterior distribution of \((f, df)\)
with the posterior expectations in bold lines, and the third row shows
the Trend Direction Index. Since the hyper-parameters, \(\bm{\Theta}\),
are known in this example, the Trend Direction Index is a deterministic
function of time. The three columns in the plot show how the posteriors
of \(f\), \(df\), and \(\mathrm{TDI}\) are updated after one, two and
four observations both forwards and backwards in time. The figure shows
how the inclusion of additional observations results in changes of the
posterior distribution of the trend and the Trend Direction Index. The
uncertainty of the forecasts remains unchanged, whereas the posterior
distributions back in time restricts the uncertainty of the curves to
accommodate the observations. The vertical dotted lines denote the point
in time after which forecasting occurs. When forecasting beyond the last
observation, the posterior of \(f\) becomes more and more dominated by
its prior implying that \(df\) becomes symmetric around zero so that the
Trend Direction Index stabilizes around \(50\%\).

\hypertarget{the-expected-trend-instability}{%
\subsection{The Expected Trend
Instability}\label{the-expected-trend-instability}}

The Expected Trend Instability was defined in Equation (\ref{eq:ETIdef})
as the expected number of roots of the trend on an interval conditional
on observed data. Now we make this concept more precise and frame it in
the context of the latent Gaussian Process model. Let \(\mathcal{I}\) be
a compact subset of the real line and consider the random càdlàg
function \begin{align*}
  N_\mathcal{I}(t) = \#\left\{s \leq t : df(s) = 0, s \in \mathcal{I}\right\}
\end{align*} counting the cumulative number of points in \(\mathcal{I}\)
up to time \(t\) where the trend is equal to zero. The Expected Trend
Instability on \(\mathcal{I}\) is equal to \begin{align*}
  \mathrm{ETI}(\mathcal{I} \mid \bm{\Theta}) = \mathop{\mathrm{E}}\left[N_{\mathcal{I}}\left(\max \mathcal{I}\right) \mid \bm{\Theta}, \mathcal{F}\right]
\end{align*} giving the expected number of zero-crossings by \(df\) on
\(\mathcal{I}\). The following proposition gives the expression for the
Expected Trend Instability under the data generating model in Equation
(\ref{eq:generatingProcess}).

\vspace{0.2cm}

\begin{proposition}
Let $\mathcal{F}$ be the $\sigma$-algebra generated by the observed data $(\mathbf{Y}, \mathbf{t})$ and $\mu_{df}$, $\mu_{d^2\!f}$, $\Sigma_{df}$, $\Sigma_{d^2\!f}$ and $\Sigma_{df,d^2\!f}$ the moments of the joint posterior distribution of $(df, d^2\!f)$ as stated in Proposition \ref{prop:GPposterior}, and assume that all assumptions A1-A4 are fulfilled. The Expected Trend Instability is then
\begin{align*}
  \mathrm{ETI}(\mathcal{I} \mid \bm{\Theta}) = \int_{\mathcal{I}} d\mathrm{ETI}(t \mid \Theta)\mathrm{d}t
\end{align*}
where $d\mathrm{ETI}$ is the local Expected Trend Instability given by
\begin{align*}
d\mathrm{ETI}(t, \mathcal{T} \mid \bm{\Theta}) = \lambda(t \mid \Theta)\phi\left(\frac{\mu_{df}(t \mid \bm{\Theta})}{\Sigma_{df}(t,t \mid \bm{\Theta})^{1/2}}\right)\left(2\phi(\zeta(t\mid \bm{\Theta})) + \zeta(t\mid \bm{\Theta})\mathop{\mathrm{Erf}}\left(\frac{\zeta(t\mid \bm{\Theta}     )}{2^{1/2}}\right)\right)
\end{align*}
and $\phi\colon\, x \mapsto 2^{-1/2}\pi^{-1/2}\exp(-\frac{1}{2}x^2)$ is the standard normal density function, $\mathop{\mathrm{Erf}}\colon\, x \mapsto 2\pi^{-1/2}\int_0^x \exp(-u^2)\mathrm{d}u$ is the error function, and $\lambda$, $\omega$ and $\zeta$ are functions defined as
\begin{align*}
  \lambda(t \mid \Theta) &= \frac{\Sigma_{d^2\!f}(t,t \mid \bm{\Theta})^{1/2}}{\Sigma_{df}(t,t \mid \bm{\Theta})^{1/2}}\left(1-\omega(t \mid \bm{\Theta})^2\right)^{1/2}\\
  \omega(t \mid \bm{\Theta}) &= \frac{\Sigma_{df,d^2\!f}(t,t \mid \bm{\Theta})}{\Sigma_{df}(t,t \mid \bm{\Theta})^{1/2}\Sigma_{d^2\!f}(t,t \mid \bm{\Theta})^{1/2}}\\
  \zeta(t\mid \bm{\Theta}) &= \frac{\mu_{df}(t\mid \bm{\Theta})\Sigma_{d^2\!f}(t,t\mid \bm{\Theta})^{1/2}\omega(t)\Sigma_{df}(t,t\mid \bm{\Theta})^{-1/2} - \mu_{d^2\!f}(t\mid \bm{\Theta})}{\Sigma_{d^2\!f}(t,t\mid \bm{\Theta})^{1/2}\left(1 - \omega(t\mid \bm{\Theta})^2\right)^{1/2}}
\end{align*}
\label{prop:ETIposterior}
\end{proposition}

\begin{figure}[htb]
\center\includegraphics{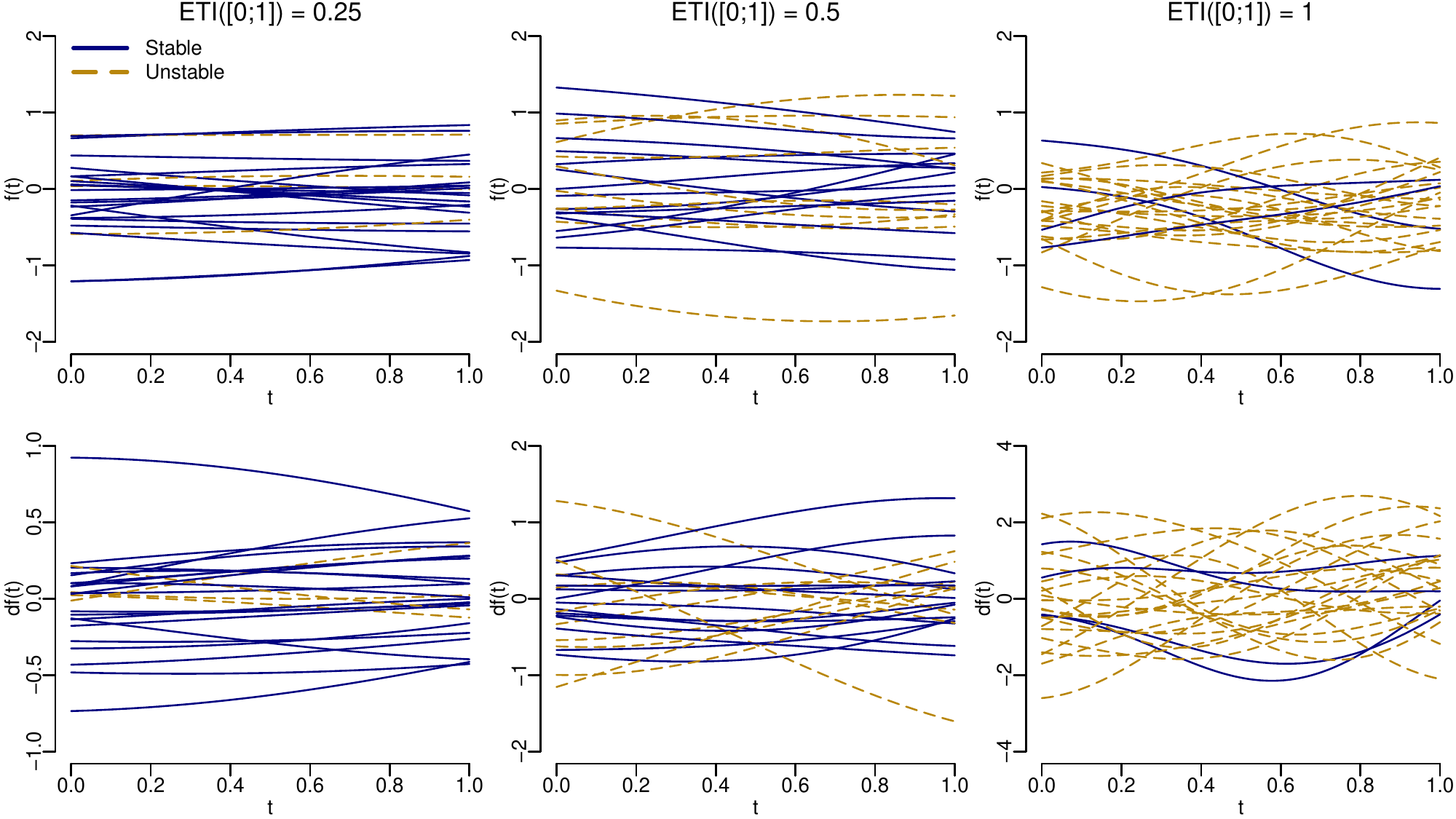}
\caption{$25$ random pairs sampled from the joint distribution of a Gaussian Process ($f$) and its derivative ($df$) with different values of Expected Trend Instability (ETI). The first row shows samples from $f$, and the second row shows samples from $df$. The columns define different values of ETI. Sample paths that are trend stable are shown by solid blue lines, and unstable sample paths are shown by dashed gold colored lines.}
\label{fig:ETIexample}
\end{figure}

To illustrate the Expected Trend Instability, Figure
\ref{fig:ETIexample} shows \(25\) random Gaussian Processes on the unit
interval simulated under three different values of \(\mathrm{ETI}\). The
sample paths are paired so that each function in the first row has an
associated trend in the second row. The different values of
\(\mathrm{ETI}\) are set to \(0.25\), \(0.5\) and \(1\) corresponding to
the expected number of times that the functions change monotonicity or
equivalently that the trend crosses zero on that interval. Sample paths
that are \textit{trend stable}, i.e., always increasing/decreasing, are
shown by solid blue lines, and sample paths that are
\textit{trend unstable}, i.e., the derivatives crosses zero, are shown
by dashed gold colored lines. It is seen that for low values of
\(\mathrm{ETI}\) most of the sample paths preserve their monotonicity on
the interval and their associated derivatives are correspondingly either
always positive or negative. For larger values of \(\mathrm{ETI}\), more
of the trends start crossing zero implying less stability in the
monotonicity of \(f\). We note that even though we are only modeling a
single curve, the Expected Trend Instability is defined in terms of a
posterior distribution of random curves which is what the figure
illustrates.

\hypertarget{sec:priorparametrization}{%
\subsection{Modeling the prior}\label{sec:priorparametrization}}

To complete the model specification in Equation
(\ref{eq:generatingProcess}) we must decide on the functional forms of
the mean and covariance functions of the Gaussian Process prior, but
there is an inherent ambiguity in how to choose these. To explain this
issue we note that a square-integrable Gaussian Process on a compact
domain can by written as
\(f(t) = \mu_{\beta}(t) + \sum_{p=1}^\infty Z_p \phi_p(t)\) where
\(Z_p\) are independent, zero mean and unit variance normally
distributed random variables, and \(\phi_p\) are pair-wise orthonormal
functions. This is known as the Karhunen--Loève representation.
Furthermore, Mercer's theorem states that any continuous covariance
function admits the representation
\(C(s,t) = \sum_{p=1}^\infty \lambda_p \phi_p(s)\phi_p(t)\) uniformly on
\(\mathcal{T} \times \mathcal{T}\) where \(\phi_p\) are the
eigen-functions of \(C\) and form an orthonormal basis of \(L^2\). This
shows that \(f\) can be written as a sum of its mean and a possibly
infinite weighted sum of functions defined implicitly through the
spectral decomposition of the covariance function. So if e.g., the
linear function is an eigen-function of the covariance function it will
be superfluous in the mean. This results in a trade-off situation where
a rich specification of the mean, \(\mu_{\beta}\), gives little space
for flexibility added by the covariance function, \(C\). On the other
hand, a spartan mean structure requires a flexible covariance structure.
The latter approach is prevalent in applied Gaussian Process regression
modeling where a very simple mean structure (often just a constant or
linear term) is used in the prior but in combination with a flexible
covariance function. We refer to Kaufman et al. (2011) for further
discussions of these issues. We note that this issue arises because our
approach focuses on a single realization of a random function. In the
case of multiple observations of \(f\) the issue would be different.

Regarding the choice of covariance function, the Squared Exponential
(SE), the Rational Quadratic (RQ), the Matern 3/2 (M3/2), and the Matern
5/2 (M5/2) are commonly used covariance functions for Gaussian Process
regression (Rasmussen and Williams 2006). These covariance functions are
given by \begin{align}
\begin{split}
  C_{\bm{\theta}}^\text{SE}(s, t) &= \alpha^2\exp\left(-\frac{(s-t)^2}{2\rho^2}\right), \quad \bm{\theta} = (\alpha, \rho) > 0\\ 
  C_{\bm{\theta}}^\text{RQ}(s, t) &= \alpha^2\left(1 + \frac{(s-t)^2}{2\rho^2\nu}\right)^{-\nu}, \quad \bm{\theta} = (\alpha, \rho, \nu) > 0\\
  C_{\bm{\theta}}^\text{M3/2}(s, t) &= \alpha^2\left(1 + \frac{\sqrt{3}\sqrt{(s-t)^2}}{\rho}\right)\exp\left(-\frac{\sqrt{3}\sqrt{(s-t)^2}}{\rho}\right), \quad \bm{\theta} = (\alpha, \rho) > 0\\
  C_{\bm{\theta}}^\text{M5/2}(s, t) &= \alpha^2\left(1 + \frac{\sqrt{5}\sqrt{(s-t)^2}}{\rho} + \frac{5(s-t)^2}{3\rho^2}\right)\exp\left(-\frac{\sqrt{5}\sqrt{(s-t)^2}}{\rho}\right), \quad \bm{\theta} = (\alpha, \rho)  > 0
\end{split}
\label{eq:covariancefunctions}
\end{align} The SE covariance function gives rise to very smooth
functions due to it being infinitely differentiable. This can be
disadvantageous in some applications as that will make it difficult for
the posterior to adapt to localized changes in smoothness. The other
listed covariance functions try to remedy this issue. Specifically, the
RQ covariance function can be derived as an infinite scale mixture of SE
covariance functions in terms of \(\rho^{-2}\). This enables the
resulting estimates to operate on different time scales simultaneously.
Smaller values of \(\nu\) will give rise to more wiggly posteriors, and
as it is one of the model's hyper-parameters its value and the implied
adaptivity will be data-driven. For \(\nu \rightarrow \infty\) the RQ
covariance function converges on \(\mathcal{T} \times \mathcal{T}\) to
the SE covariance function. The Matern covariance functions shown here
explicitly for a generally continuous parameter \(\nu = 3/2\) and
\(\nu = 5/2\) belong to a larger class of covariance functions. The
general expression is involved and includes a modified Bessel function
for which derivatives are difficult to calculate. In this class of
covariance functions \(\nu\) directly controls the differentiability
which again can be chosen in a data-driven manner.

If non-stationary components are suspected to be presented in the data
generating process, these can be included in the covariance structure as
part of defining the prior. One example is the non-stationary covariance
function for periodic components discussed by MacKay (1998). Another
noteworthy example is the neural network covariance structure of
Williams (1998). In fact, Neal (2012) showed that a Bayesian neural
network with one hidden layer converges to a Gaussian Process when the
number of hidden neurons goes to infinity. The actual form of covariance
function implied by the neural network depends on the prior for the
network weights and the activation functions.

It should be noted that sums and products of covariance functions
produce valid covariance functions. Thus, a very wide range of flexible
covariance structures can be constructed from simpler building blocks.

We conclude this section by noting two covariance functions that cannot
be used in our model. This is because they are both in violation with
Assumption A3 which can be verified by a straight-forward calculation.
The first case is the M3/2 covariance function in Equation
(\ref{eq:covariancefunctions}). While it can be used to estimate the
Trend Direction Index, it cannot be used if an estimate of the Expected
Trend Instability is also required. Another example is
\(C_{\bm{\theta}}^\text{OU}(s, t) = \alpha^2 \exp\left(-|s-t|/\rho\right)\).
This is the covariance function of the Ornstein-Uhlenbeck process which
is mean-squared continuous but not mean-squared differentiable. This
also fails to satisfy the required assumption.

\hypertarget{sec:gaussianassumptions}{%
\subsection{The Gaussian assumptions and non-normal
outcomes}\label{sec:gaussianassumptions}}

The closed-form expressions derived above depend on the assumptions of
normality on the latent scale and for the conditional distribution of
the observed data. For outcomes that are conditionally non-normal the
posterior distributions are in general analytically intractable. In some
cases one can see our assumptions as an approximation facilitated by the
central limit theorem e.g., in the case where the outcome is a
proportion and the number of experiments is large, or if the outcome is
count data and the rate is not too low. The conditional variance,
\(\sigma^2\) in Equation (\ref{eq:generatingProcess}), can therefore be
changed to reflect the structure of the variance of such limiting
distribution.

One possibility for altering the model is to retain the structure but
apply a transformation of the observed data. This amounts to altering
the second part of Equation (\ref{eq:generatingProcess}) to
\begin{align*}
  g(Y_i) \mid t_i, f(t_i), \bm{\Theta} &\overset{iid}{\sim} N(f(t_i), \sigma^2)
\end{align*} where \(g\) is a known, monotone function. Our model could
directly be fitted to the transformed outcomes, but the posterior
estimates will also be on the transformed scale and may therefore be
difficult to interpret.

Another possibility is to consider both the outcome and the latent
function on the same transformed scale. This alters Equation
(\ref{eq:generatingProcess}) to \begin{align*}
\begin{split}
  g(f) \mid \bm{\beta}, \bm{\theta} &\sim \mathcal{GP}(\mu_{\bm{\beta}}(\cdot), C_{\bm{\theta}}(\cdot,\cdot))\\
  g(Y_i) \mid t_i, g(f(t_i)), \bm{\Theta} &\overset{iid}{\sim} N(g(f(t_i)), \sigma^2)
\end{split}
\end{align*} Our model is again directly applicable under this
alternative data generating model. It follows that the trend on this
scale is \(g(f(t))' = f'(t)g'(f(t))\), and the Trend Direction Index
will therefore be equal to
\(\mathrm{TDI}(t, \delta \mid \bm{\Theta}) = P(f'(t + \delta)g'(f(t + \delta)) > 0 \mid g(\mathbf{Y}), \mathbf{t}, \bm{\Theta})\).
This can, however, be transformed back to the original scale of \(df\)
by normalization since we can sample from the posterior distribution of
\(g(f(t))\) by the Gaussian assumptions on the transformed scale.
Therefore, it is also possible to obtain samples from \(g'(f(t))\) as
\(g^{-1}\) and \(g'\) are known. The Trend Direction Index on the
original scale can in this way be obtained by a Monte Carlo
approximation using the transformed samples.

\hypertarget{sec:estimation}{%
\section{Estimation}\label{sec:estimation}}

The trend indices depend on the hyper-parameters, \(\bm{\Theta}\), of
the latent Gaussian Process. These must be estimated from the observed
data, and we consider two different estimation procedures: maximum
likelihood estimation and a fully Bayesian estimator.

The difference between the maximum likelihood and the Bayesian method is
that they give rise to two different interpretations of the Trend
Direction Index. In the former, the index is a deterministic function,
while in the latter it is a random function governed by the posterior
distribution of \(\bm{\Theta}\). The maximum likelihood estimator is
also known as the empirical Bayes approach since the hyper-parameters
are estimated from data using the marginal likelihood of the model. This
can be seen as an approximation to the Bayesian model where the
hyper-parameters are fixed at their most likely values instead of being
integrated out.

The maximum likelihood estimator consists of finding the values of the
hyper-parameters that maximize the marginal likelihood of the observed
data and plugging these into the expressions of the posterior
distributions and the trend indices. The marginal distribution of
\(\mathbf{Y}\) can be found by integrating out the distribution of the
latent Gaussian Process in the conditional specification
\(\mathbf{Y} \mid f(\mathbf{t}), \mathbf{t}, \bm{\Theta}\) in Equation
(\ref{eq:generatingProcess}). Since the observation model consists of
normal distributed random variables conditional on the latent Gaussian
Process, the marginal distribution is multivariate normal with
expectation \(\mu_{\bm{\beta}}(\mathbf{t})\) and \(n \times n\)
covariance matrix
\(C_{\bm{\theta}}(\mathbf{t}, \mathbf{t}) + \sigma^2 I\). The marginal
log-likelihood function is therefore \begin{align}
\log L(\bm{\Theta} \mid \mathbf{Y}, \mathbf{t}) &\propto - \frac{1}{2}\log |C_{\bm{\theta}}(\mathbf{t}, \mathbf{t}) + \sigma^2 I| - \frac{1}{2}(\mathbf{Y} - \mu_{\bm{\beta}}(\mathbf{t}))^T\left[C_{\bm{\theta}}(\mathbf{t}, \mathbf{t}) + \sigma^2 I\right]^{-1}(\mathbf{Y} - \mu_{\bm{\beta}}(\mathbf{t}))\label{eq:margloglik}
\end{align} and the maximum likelihood estimate
\(\widehat{\bm{\Theta}^\text{ML}} = \mathop{\mathrm{arg\,sup}}_{\bm{\Theta} = (\bm{\beta}, \bm{\theta}, \sigma)} \log L(\bm{\Theta} \mid \mathbf{Y}, \mathbf{t})\)
can be obtained by numerical optimization or found as the roots to the
score equations
\(\nabla_{\bm{\Theta}} \log L(\bm{\Theta} \mid \mathbf{Y}, \mathbf{t}) = 0\).
This estimate can be plugged in to the expressions for the posterior
distributions of \((f, df, d^2\!f)\) in Proposition
\ref{prop:GPposterior} enabling simulation of the posterior
distributions at any vector of time points. Estimates of the Trend
Direction Index and the Expected Trend Instability are similarly
\(\mathrm{TDI}(t, \delta \mid \widehat{\bm{\Theta}^\text{ML}})\) and
\(\mathrm{ETI}(\mathcal{I} \mid \widehat{\bm{\Theta}^\text{ML}})\)
according to Propositions \ref{prop:TDIposterior} and
\ref{prop:ETIposterior} respectively, and the predictive distribution of
a new observation is equal to Equation (\ref{eq:PPD}) with the plug-in
estimate \(\bm{\Theta} = \widehat{\bm{\Theta}^\text{ML}}\).

The maximum likelihood estimator is easy to implement and fast to
compute, but it is difficult to propagate the uncertainties of the
parameter estimates through to the posterior distributions and the trend
indices. This is disadvantageous since in order to conduct a qualified
assessment of trendiness we are not only interested in point estimates
but also the associated uncertainties. A Bayesian estimator, while
slower to compute, facilitates a straightforward way to encompass all
uncertainties in the final estimates. To specify a Bayesian estimator we
must augment the data generating model in Equation
(\ref{eq:generatingProcess}) with another hierarchical level specifying
the prior distribution of the hyper-parameters. We therefore add the
following level \begin{align*}
  (\bm{\beta}, \bm{\theta}, \sigma) \sim G(\bm{\Theta} \mid \bm{\Psi}, \mathbf{t})
\end{align*} to the specification where \(G\) is some family of
distribution indexed by a vector \(\bm{\Psi}\). The joint distribution
of the model can be factorized as \begin{align*}
  P(\mathbf{Y}, f(\mathbf{t}), \bm{\Theta} \mid \bm{\Psi}, \mathbf{t}) = P(\mathbf{Y} \mid f(\mathbf{t}), \bm{\Theta}, \bm{\Psi}, \mathbf{t})P(f(\mathbf{t}) \mid \bm{\Theta}, \bm{\Psi}, \mathbf{t})G(\bm{\Theta} \mid \bm{\Psi}, \mathbf{t})
\end{align*} and each conditional probability is specified by a
sub-model in the augmented hierarchy. We always condition on
\(\bm{\Psi}\) and \(\mathbf{t}\) as they are considered fixed. The
posterior distribution of the hyper-parameters given the observed data
is then \begin{align}
\begin{split}
  P(\bm{\Theta} \mid \mathbf{Y}, \bm{\Psi}, \mathbf{t}) & = \frac{G(\bm{\Theta} \mid \bm{\Psi}, \mathbf{t})P(\mathbf{Y} \mid \bm{\Theta}, \bm{\Psi}, \mathbf{t})}{P(\mathbf{Y} \mid \bm{\Psi}, \mathbf{t})}\\
   &= \frac{G(\bm{\Theta} \mid \bm{\Psi}, \mathbf{t}) \int P(\mathbf{Y} \mid f(\mathbf{t}), \bm{\Theta}, \bm{\Psi}, \mathbf{t})dP(f(\mathbf{t}) \mid \bm{\Theta}, \bm{\Psi}, \mathbf{t})}{\iint P(\mathbf{Y} \mid f(\mathbf{t}), \bm{\Theta}, \bm{\Psi}, \mathbf{t})dP(f(\mathbf{t}) \mid \bm{\Theta}, \bm{\Psi}, \mathbf{t})dG(\bm{\Theta} \mid \bm{\Psi}, \mathbf{t})}
\end{split}
\label{eq:posteriorHyper}
\end{align} and we let
\(\widetilde{\bm{\Theta}} \sim P(\bm{\Theta} \mid \mathbf{Y}, \bm{\Psi}, \mathbf{t})\).
The posterior distribution of \(\bm{\Theta}\) induces corresponding
distributions over the trend indices according to
\(\mathrm{TDI}(t, \delta \mid \widetilde{\bm{\Theta}})\),
\(d\mathrm{EDI}(t, \mathcal{T}, \mid \widetilde{\bm{\Theta}})\) and
\(\mathrm{EDI}(\mathcal{I} \mid \widetilde{\bm{\Theta}})\). For example,
the Trend Direction Index in the Bayesian formulation is a surface in
\((t, \delta)\) where each value is a distribution over probabilities.
We suggest to summarize the trend indices by their posterior quantiles.
For the Trend Direction Index we summarize its posterior distribution by
functions \(Q_\tau\) such that \begin{align*}
  P\left(\mathrm{TDI}\left(t, \delta \mid \widetilde{\bm{\Theta}}\right) \leq \tau\right) = Q_\tau(t, \delta)
\end{align*} with for example
\(\tau \in \left\{0.025, 0.5, 0.975\right\}\) to obtain 95\% credible
intervals. In the Bayesian model the predictive distribution in Equation
(\ref{eq:PPD}) should be averaged across the posterior distribution of
the hyper-parameters. This leads to the posterior predictive
distribution \begin{align*}
Y^\ast(t^\ast) \mid t^\ast, \mathbf{Y}, \mathbf{t}, \bm{\Psi} = \int_{\bm{\Theta}} P(Y^\ast(t^\ast) \mid t^\ast, \mathbf{Y}, \mathbf{t}, \bm{\Theta}, \bm{\Psi})dP(\bm{\Theta} \mid \mathbf{Y}, \bm{\Psi}, \mathbf{t})
\end{align*} where the integral is in practice approximated through the
MCMC samples.

We have implemented both the maximum likelihood and the Bayesian
estimator in Stan (Carpenter et al. 2017) and R (R Core Team 2018) in
combination with the \texttt{rstan} package (Stan Development Team
2018). Stan is a probabilistic programming language enabling fully
Bayesian inference using Markov chain Monte Carlo sampling. The Stan
implementation of the maximum likelihood estimator requires the marginal
maximum likelihood estimates of the parameters supplied as data, and
from these it will simulate random realizations of the posterior
distribution of \((f, df, d^2\!f)\) on a user-supplied grid of time
points and return point estimates of \(\mathrm{TDI}\) and
\(d\mathrm{ETI}\). The latter can then be integrated numerically to
obtain the Expected Trend Instability on an interval. The Bayesian
estimator requires a specification of \(\bm{\Psi}\) supplied as data and
from that it will generate realizations of
\((\widetilde{\bm{\Theta}}_1, \ldots, \widetilde{\bm{\Theta}}_K)\) from
Equation (\ref{eq:posteriorHyper}). These samples are then used to
obtain the posterior distribution of the Trend Direction Index by
\((\mathrm{TDI}(t, \delta \mid \widetilde{\bm{\Theta}}_1), \ldots, \mathrm{TDI}(t, \delta \mid \widetilde{\bm{\Theta}}_K))\)
and similarly for \(d\mathrm{ETI}\) according to Propositions
\ref{prop:TDIposterior} and \ref{prop:ETIposterior}.

\hypertarget{sec:modelselection}{%
\subsection{Model selection}\label{sec:modelselection}}

In connection with Section \ref{sec:priorparametrization} on choosing
the mean and covariance structure for the Gaussian Process prior we
propose a practical approach based on specifying a set of candidate
models and choosing the best fitting model among these according to a
cross-validation procedure. Different types of cross-validation can be
performed depending on the purpose of the analysis. In some cases one
stands at the end of the data collection and wants to look at what has
been observed. In this cases it would be natural to condition on all the
observed data and perform leave-one-out cross-validation. In other cases
one is interested in forecasting the trend and the associated indices,
and here it would be natural to take the direction of time into account.
This could be done by e.g., performing one-step-ahead cross-validation
in which the observed data are compared to a sequence of models
forecasting one step ahead in time based on successive partitioning of
the time series. For more information on such procedures see e.g.,
Bergmeir and Benítez (2012).

To perform the leave-one-out cross-validation we consider a set of
candidate models indexed by \(\mathcal{M}\). This set would typically
include different parameterizations of the mean and covariance function
of \(f\). Turn by turn, a single data pair \((Y_i, t_i)\) is excluded,
and we let the corresponding leave-one-out data be denoted
\((\mathbf{Y}_{-i}, \mathbf{t}_{-i})\). Based on the leave-one-out data
the hyper-parameters are estimated for each model by maximizing the
marginal log-likelihood in Equation (\ref{eq:margloglik}) and given by
\begin{align*}
\widehat{\bm{\Theta}}_{-i}^\mathcal{M} = \mathop{\mathrm{arg\,sup}}_{\bm{\Theta}} \log L(\bm{\Theta} \mid \mathbf{Y}_{-i}, \mathbf{t}_{-i}) 
\end{align*} Plugging these estimates into the expression for the
posterior expectation of \(f\) in Proposition \ref{prop:GPposterior} we
obtain the leave-one-out predictions, and the mean squared error of
prediction (or another loss function) can be calculated by comparing the
predictions and the observed values averaged across all data points as
\begin{align*}
  \text{MSPE}_{\text{LOO}}^\mathcal{M} = \frac{1}{n}\sum_{i=1}^{n} \left(Y_i - \mathop{\mathrm{E}}[f(t_i) \mid \mathbf{Y}_{-i}, \mathbf{t}_{-i}, \widehat{\bm{\Theta}}_{-i}^\mathcal{M}]\right)^2
\end{align*} The selected model among the candidate set is
\(\mathcal{M}_{\text{opt}} = \mathop{\mathrm{arg\,min}}_\mathcal{M} \text{MSPE}_{\text{LOO}}^\mathcal{M}\).
Different cross-validation schemes can be performed in a similar manner
by modifying how the leave-one-out data sets are constructed.

We note that with our model being implemented in Stan, efficient
approximate leave-one-out cross-validation and model comparison using
the LOOIC criterion can be directly performed with the \texttt{loo}
package (Vehtari et al. 2019).

\hypertarget{sec:simulation}{%
\section{Simulation study}\label{sec:simulation}}

To assess the performance of our method we performed a simulation study.
We generated \(r = 1,\ldots, 10,000\) random Gaussian Processes on the
unit interval with zero mean and the Squared Exponential (SE) covariance
function (see Equation (\ref{eq:covariancefunctions})) with parameters
\(\alpha = 1\) and \(\rho = \frac{\sqrt{3}}{{2\pi}}\) in 15 different
scenarios in which we varied the number of observation points
(\(n = 25, 50, 100\)) and the measurement noise
(\(\sigma = 0.025, 0.05, 0.1, 0.15, 0.2\)). The Supplementary Material
shows 50 random sample paths for each scenario.

In each of the \(r\) simulations we know the true latent functions
\((f_r, df_r)\), and by fitting our model we obtain estimates
\(\widehat{f_r^\text{GP}}\) and \(\widehat{df_r^\text{GP}}\)
corresponding to the posterior expectations in Proposition
\ref{prop:GPposterior}, and \(\mathrm{TDI}_r\) and \(d\mathrm{ETI}_r\)
from Propositions \ref{prop:TDIposterior} and \ref{prop:ETIposterior}.
We compare these estimates to the truths using two different measures:
an integrated residual and the squared \(L^2\) norm. The integrated
residuals are defined as
\(\int_0^1 (f_r(t) - \widehat{f_r^\text{GP}}(t))\mathrm{d}t\) and
similarly for \(\widehat{df_r^\text{GP}}\). For the Trend Direction
Index the cumulative residual is defined as
\(\int_0^1 (1(df_r(t) > 0) - \mathrm{TDI}_r(t))\mathrm{d}t\) where \(1\)
denotes the indicator function. For the Expected Trend Instability the
cumulative residual is defined as
\(\int_0^1(N_r(t) - d\mathrm{ETI}_r(t))\mathrm{d}t\) where \(N_r(t)\) is
the càdlàg counting process that jumps with a value of 1 every time
\(df_r\) has a root on the interval. If our estimates are unbiased we
expect these integrated residuals to have zero mean across the
simulations. The squared \(L^2\) norms are defined in a similar manner
for all the quantities as e.g.,
\(\int_0^1 (f_r(t) - \widehat{f_r^\text{GP}}(t))^2\mathrm{d}t\) and
reflect the variability of the estimates.

For comparison we employed the Trend Filtering method implemented in the
R package \texttt{genlasso} (Arnold and Tibshirani 2019) on the same
simulated data and reported similar measures for its estimated mean,
\(\widehat{f^\text{TF}}\), and derivative, \(\widehat{df^\text{TF}}\),
using 10-fold cross-validation of the penalty parameter. We only compare
the estimates of the latent mean and its derivative between the two
approaches as Trend Filtering does not provide a probability
distribution for the derivative.

Summary statistics from the simulation study are shown in Table
\ref{tab:simulationSummary}. It is seen that both our model and Trend
Filtering provide unbiased estimates in all scenarios. Looking at the
squared \(L^2\) norm our estimates of \(f\) and \(df\) show very small
variability across all scenarios, while the estimates from Trend
Filtering showed an increase in variability for increasing measurement
noise and a decrease for increasing number of observations. This is
expected as we simulate random functions with continuous sample paths,
and Trend Filtering estimates piece-wise linear approximations. While
this leads to unbiased estimates, the variability of these estimates are
larger, and this is more pronounced for smaller number of observations.

The variability of the Trend Direction Index and the Expected Trend
Instability as measured by the squared \(L^2\) norm were low in all
scenarios but increased with the magnitude of the measurement noise as
expected. In a few cases the estimated Expected Trend Instability was
far away from its true value. This was especially pronounced in
scenarios with a small number of observations and a high measurement
noise. The reason was that \(\widehat{f^\text{GP}}\) degenerated to
either a constant function or to a perfect interpolation of the observed
data. This is a consequence of the hyper-parameters being weakly
identified under such circumstances. These cases are still part of the
reported summary statistics in Table \ref{tab:simulationSummary}. For
such cases, the model fit can be regularized through the priors of the
hyper-parameters, but this must be determined on a case-by-case basis.

\begin{table}[!h]

\caption{\label{tab:unnamed-chunk-1}\label{tab:simulationSummary}Summary statistics from the simulation study. Each value is the mean across 10,000 simulations except for ETI where the median is reported. Superscript GP denotes our proposed method and TF denotes Trend Filtering. Numbers have been rounded to three decimal places.}
\centering
\fontsize{9}{11}\selectfont
\begin{tabular}[t]{rr|>{}rrrrrr|>{}rrrrrr}
\toprule
\multicolumn{2}{c}{ } & \multicolumn{6}{c}{Integrated residual} & \multicolumn{6}{c}{Squared L2 norm} \\
n & $\sigma$ & $\widehat{f^\text{GP}}$ & $\widehat{df^\text{GP}}$ & $\widehat{f^\text{TF}}$ & $\widehat{df^\text{TF}}$ & TDI & ETI & $\widehat{f^\text{GP}}$ & $\widehat{df^\text{GP}}$ & $\widehat{f^\text{TF}}$ & $\widehat{df^\text{TF}}$ & TDI & ETI\\
\midrule
25 & 0.025 & 0 & 0 & 0.000 & 0.000 & 0.000 & 0.000 & 0.000 & 0.000 & 0.041 & 0.504 & 0.011 & 0.008\\
25 & 0.050 & 0 & 0 & 0.000 & 0.000 & 0.000 & -0.002 & 0.001 & 0.001 & 0.130 & 1.188 & 0.021 & 0.018\\
25 & 0.100 & 0 & 0 & 0.002 & 0.003 & 0.001 & -0.008 & 0.003 & 0.004 & 0.408 & 2.988 & 0.037 & 0.040\\
25 & 0.150 & 0 & 0 & -0.001 & -0.002 & -0.001 & -0.020 & 0.005 & 0.008 & 0.818 & 5.248 & 0.051 & 0.064\\
25 & 0.200 & 0 & 0 & -0.003 & -0.003 & -0.001 & -0.034 & 0.009 & 0.014 & 1.343 & 8.248 & 0.063 & 0.094\\
\hline
50 & 0.025 & 0 & 0 & 0.000 & 0.000 & 0.000 & 0.000 & 0.000 & 0.000 & 0.026 & 0.535 & 0.009 & 0.006\\
50 & 0.050 & 0 & 0 & 0.000 & 0.000 & 0.000 & -0.002 & 0.000 & 0.001 & 0.081 & 1.287 & 0.016 & 0.012\\
50 & 0.100 & 0 & 0 & 0.000 & 0.000 & 0.000 & -0.006 & 0.001 & 0.002 & 0.244 & 3.348 & 0.028 & 0.028\\
50 & 0.150 & 0 & 0 & 0.000 & 0.001 & 0.000 & -0.012 & 0.003 & 0.005 & 0.469 & 6.353 & 0.038 & 0.045\\
50 & 0.200 & 0 & 0 & 0.000 & 0.003 & 0.001 & -0.020 & 0.005 & 0.008 & 0.759 & 9.832 & 0.050 & 0.068\\
\hline
100 & 0.025 & 0 & 0 & 0.000 & 0.000 & 0.000 & 0.000 & 0.000 & 0.000 & 0.016 & 0.399 & 0.007 & 0.004\\
100 & 0.050 & 0 & 0 & 0.000 & 0.001 & 0.000 & -0.001 & 0.000 & 0.000 & 0.050 & 1.020 & 0.012 & 0.009\\
100 & 0.100 & 0 & 0 & 0.000 & 0.000 & 0.000 & -0.003 & 0.001 & 0.001 & 0.150 & 2.187 & 0.022 & 0.019\\
100 & 0.150 & 0 & 0 & 0.000 & -0.003 & 0.000 & -0.007 & 0.002 & 0.002 & 0.283 & 4.454 & 0.030 & 0.031\\
100 & 0.200 & 0 & 0 & 0.002 & 0.002 & 0.001 & -0.012 & 0.003 & 0.004 & 0.450 & 6.140 & 0.038 & 0.045\\
\bottomrule
\end{tabular}
\end{table}

\hypertarget{sec:application}{%
\section{Applications}\label{sec:application}}

\hypertarget{trend-of-proportion-of-danish-smokers}{%
\subsection*{Trend of proportion of Danish
smokers}\label{trend-of-proportion-of-danish-smokers}}
\addcontentsline{toc}{subsection}{Trend of proportion of Danish smokers}

A report published by The Danish Health Authority in January 2019
updated the estimated proportion of daily or occasional smokers in
Denmark with new data from 2018 (The Danish Health Authority 2019). The
data was based on an online survey including 5017 participants. The
report also included data on the proportion of smokers in Denmark during
the last 20 years which was shown in Figure \ref{fig:rawDataPlot}. The
report was picked up by several news papers under headlines stating that
the proportion of smokers in Denmark had significantly increased for the
first time in two decades (Navne, Schmidt, and Rasmussen 2019). The
report published no statistical analyses for this statement but wrote,
that because the study population is so large, then more or less all
differences become statistically significant at the \(5\%\) level (this
was written as a \(95\%\) significance level in the report).

This data set provides an instrumental way of exemplifying our two
proposed measures of trendiness. In this application we wish to assess
the statistical properties of the following questions:

\begin{itemize}
\item[Q1:]{Is the proportion of smokers increasing in the year 2018 conditional on data from the last 20 years?}
\item[Q2:]{If the proportion of smokers is currently increasing, when did this increase probably start?}
\item[Q3:]{Is it the first time during the last 20 years that the trend in the proportion of smokers has changed?}
\end{itemize}

A naive approach for trying to answer questions Q1 and Q2 is to apply
sequential \(\chi^2\)-tests in a \(2\times 2\) table. Table
\ref{tab:chisqtests} shows the p-values for the \(\chi^2\)-test of
independence between the proportion of smokers in 2018 and each of the
five previous years. Using a significance level of \(5\%\) the
conclusion is ambiguous. Compared to the previous year, there was no
significant change in the proportion in 2018. Three out of these five
comparisons fail to show a significant change in proportions. It is
therefore evident that such point-wise testing is not sufficiently
perceptive to catch the underlying continuous development.

\begin{table}[htbp]
\center
\begin{tabular}{c|rrrrr}
 & 2017 & 2016 & 2015 & 2014 & 2013\\ \hline 
p-value & 0.074 & \textbf{0.020} & 0.495 & \textbf{0.012} & 0.576
\end{tabular}
\caption{p-values obtained from $\chi^2$-tests of independence between the proportion of smokers in 2018 and the five previous years. Numbers in bold are statistically significant differences at the $5\%$ level.}
\label{tab:chisqtests}
\end{table}

Similarly, a simple approach for trying to answer question Q3 would be
to look at the cumulative number of times that the difference in
proportion between consecutive years changes sign. In the data set there
were nine changes in the sign of the difference between the proportion
in each year and the proportion in the previous year giving this very
crude estimate of the number of times that the trend has changed. This
approach suffers from the facts that it is based on a finite difference
approximation at the sampling points to the continuous derivative, and
that it uses the noisy measurements instead the latent function.
Consequently, the changes in trend are quite unstable.

We now present an analysis of the data set using our method. As a
specification of the latent function we considered three different mean
functions, a constant mean \(\mu_{\bm{\beta}}(t) = \beta_0\), a linear
mean \(\mu_{\bm{\beta}}(t) = \beta_0 + \beta_1 t\), and a quadratic mean
\(\mu_{\bm{\beta}}(t) = \beta_0 + \beta_1 t + \beta_2 t^2\) and the four
different covariance functions given in Equation
(\ref{eq:covariancefunctions}). This gives a total of 12 different
candidate models to compare. Since we condition on all the observed data
and are not interested in forecasting in this application we performed
the model comparison by leave-one-out cross-validation as discussed in
Section \ref{sec:modelselection}.

Table \ref{tab:looTale} shows the mean squared error of prediction for
each candidate model. For the models with a Rational Quadratic
covariance function and a linear and a quadratic mean function the
parameter \(\nu\) diverged numerically implying convergence to the
Squared Exponential covariance function. Comparing the leave-one-out
mean squared error of prediction, the prior distribution of \(f\) in the
optimal model has a constant mean function and a Rational Quadratic
covariance function. The marginal maximum likelihood estimates of the
parameters in the optimal model were \begin{align}
  \widehat{\beta_0^\text{ML}} = 28.001, \quad \widehat{\alpha^\text{ML}} = 4.543, \quad \widehat{\rho^\text{ML}} = 4.438, \quad \widehat{\nu^\text{ML}} = 1.020, \quad \widehat{\sigma^\text{ML}} = 0.622\label{eq:mlEstimates}
\end{align}

\begin{table}[htbp]
\center
\begin{tabular}{l|rrrr}
 & SE & RQ & Matern 3/2 & Matern 5/2\\ \hline
Constant & 0.682 & \textbf{0.651} & 0.687 & 0.660\\
Linear & 0.806 & $\Leftarrow    $ & 0.896 & 0.865\\
Quadratic & 0.736 & $\Leftarrow$ & 0.800 & 0.785
\end{tabular}
\caption{Leave-one-out cross-validated mean squared error of prediction for each of the 12 candidate models. $\Leftarrow$ indicates numerical convergence to the SE covariance function.}
\label{tab:looTale}
\end{table}

\begin{figure}[htb]
\center\includegraphics{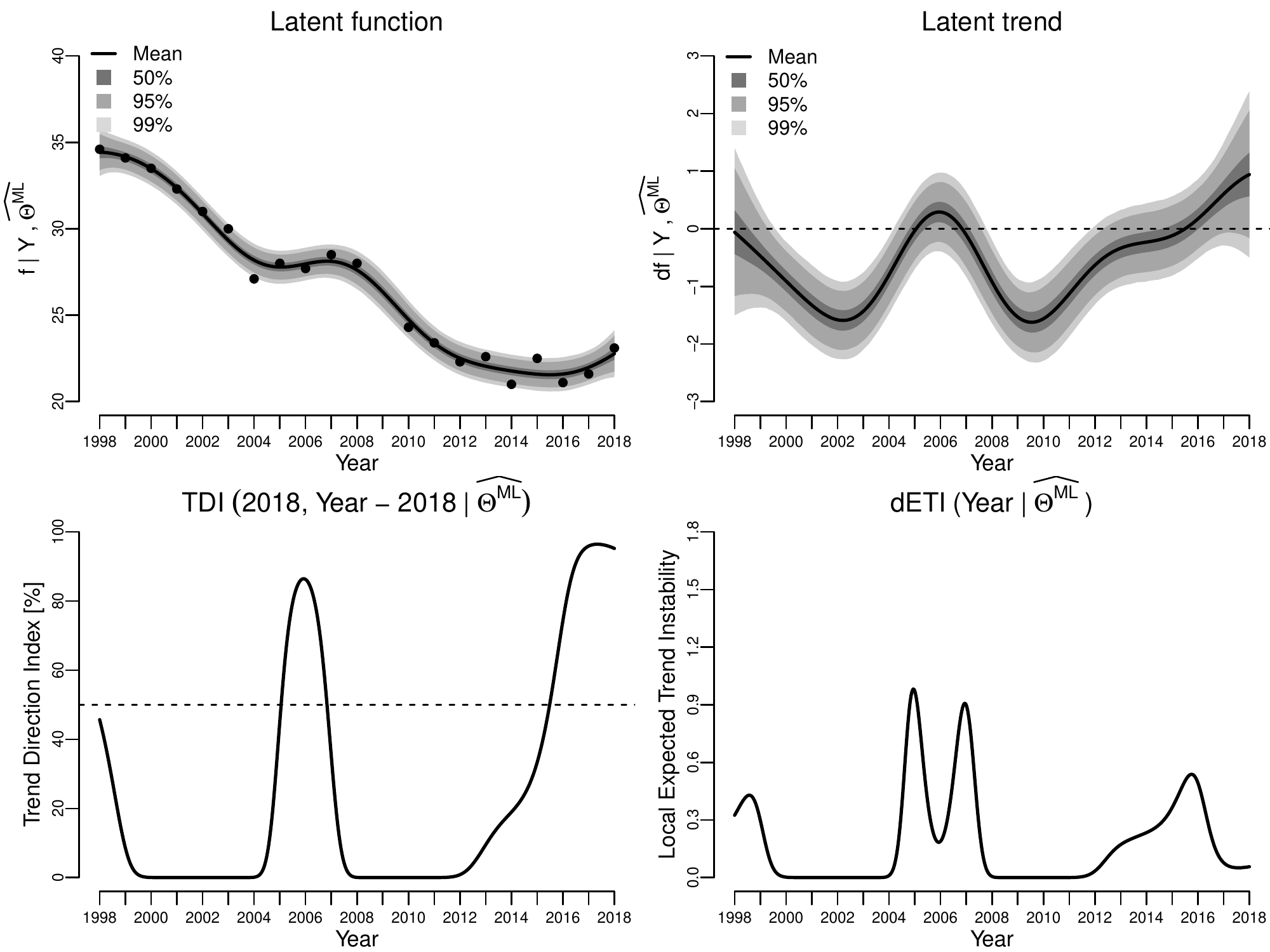}
\caption{Results from fitting the latent Gaussian Process model by maximum likelihood. The first row shows the posterior distributions of $f$ (left) and $df$ (right) with the posterior means in bold and gray areas showing point-wise probability intervals for the posterior distribution. The second row shows the estimated Trend Direction Index (left) and the local Expected Trend Instability (right).}
\label{fig:likFitPlot}
\end{figure}

Figure \ref{fig:likFitPlot} shows the fit of the model by the maximum
likelihood method. The plots were obtained by plugging the maximum
likelihood estimates into the expressions for the posterior
distributions of \(f\) and \(df\) defined in Proposition
\ref{prop:GPposterior}, the Trend Direction Index in Proposition
\ref{prop:TDIposterior}, and the Expected Trend Instability in
Proposition \ref{prop:ETIposterior}. The predictions were performed on
an equidistant grid of \(500\) time points spanning the 20 years. The
plot of the posterior trend (top right) shows two regions in time where
the posterior mean of the derivative is positive, one around
\([2004; 2008]\) and one shortly after \(2015\) and until the end of the
observation period. The Trend Direction Index (bottom left) quantifies
this positive trendiness as a probability standing in \(2018\) and
looking back in time while also taking the uncertainty into account. The
bottom right panel shows the local Expected Trend Instability and its
integral is to the expected number of times that the trend has changed
sign. Table \ref{tab:summaries} summarizes the maximum likelihood
estimates of the Trend Direction Index at the end of the observation
period and the previous five years as well as the Expected Trend
Instability during the full observation period and during only the last
ten years. \(\text{Crosspoint}\) is the first point in time during the
last ten years where the Trend Direction Index became greater than
\(50\%\) i.e., \begin{align*}
\text{Crosspoint} = \mathop{\mathrm{arg\,min}}_{t \in [-10; 0]} \left\{2018 + t : \text{TDI}(2018, 2018 - t) \geq 50\%\right\}
\end{align*}

Based on the results from the maximum likelihood analysis we may answer
the questions by stating that the expected proportion of smokers in
Denmark is currently increasing with a probability of \(95.24\%\). This
is, however, not a recent development as the probability of an
increasing proportion has been greater than \(50\%\) since the middle of
2015. This can be compared to the sequential \(\chi^2\)-tests in Table
\ref{tab:chisqtests}, which gave a cruder and less consistent result.
The estimated values of \(\mathrm{ETI}\) in Table \ref{tab:summaries}
show that there has been an average of \(3.68\) changes in the
monotonicity of the proportion during the last 20 years. This value does
support the statement by the news outlets that it is the first time in
20 years that the trend has changed. The value, however, reduces to 1.39
when only looking 10 years back, which is slightly less than half the
ETI for the longer period.

We also applied the Bayesian estimator to the data using the same prior
mean and covariance structure. The Bayesian estimator requires a prior
distribution of the hyper-parameters. We used independent priors of the
form \begin{align*}
G(\bm{\Theta} \mid \bm{\Psi}, \mathbf{t}) = G(\beta_0 \mid \Psi_{\beta_0}, \mathbf{t})G(\alpha \mid \Psi_{\alpha}, \mathbf{t})G(\rho \mid \Psi_{\rho}, \mathbf{t})G(\nu \mid \Psi_{\nu}, \mathbf{t})G(\sigma \mid \Psi_{\sigma}, \mathbf{t})
\end{align*} where each prior was a heavy-tailed distribution with a
moderate variance centered at the maximum likelihood estimates. We used
the following distributions \begin{alignat*}{3}
 \beta_0 &\sim T\left(\widehat{\beta_0^\text{ML}}, 3, 3\right), &\quad \alpha &\sim \text{Half-}T\left(\widehat{\alpha^\text{ML}}, 3, 3\right), &\quad \rho &\sim \text{Half-}N\left(\widehat{\rho^\text{ML}}, 1\right)\\   
 \nu &\sim \text{Half-}T\left(\widehat{\nu^\text{ML}}, 3, 3\right), & \sigma &\sim \text{Half-}T\left(\widehat{\sigma^\text{ML}}, 3, 3\right)  & 
\end{alignat*} where the maximum likelihood values are given in Equation
(\ref{eq:mlEstimates}) and \(\text{Half-}T(\cdot, \cdot, \mathrm{df})\)
and \(\text{Half-}N(\cdot, \cdot)\) denotes the location-scale half T-
and normal distribution functions with \(\mathrm{df}\) degrees of
freedom due to the requirement of positivity. We ran four independent
Markov chains for 25,000 iterations each with half of the iterations
used for warm-up and discarded. Convergence was assessed by trace plots
of the MCMC draws and the potential scale reduction factor,
\(\widehat{R}\), of Gelman and Rubin (1992). The trace plots are
included in the Supplementary Material.

Figure \ref{fig:bayesFitPlot} shows the results from the Bayesian
estimator. In this model both trend indices are time-varying posterior
distributions and the top row of the shows the posterior distributions
of the Trend Direction Index (left) and the Local Expected Trend
Instability (right) summarized by time-dependent quantiles. The bottom
row shows posterior density estimates of the Expected Trend Instability
during the last twenty years (left) and during the last ten years
(right). The same summary statistics as for the maximum likelihood
analysis are given in Table \ref{tab:summaries} but here stated in terms
of posterior medians and \(2.5\%\) and \(97.5\%\) posterior quantiles.

\begin{figure}[htb]
\center\includegraphics{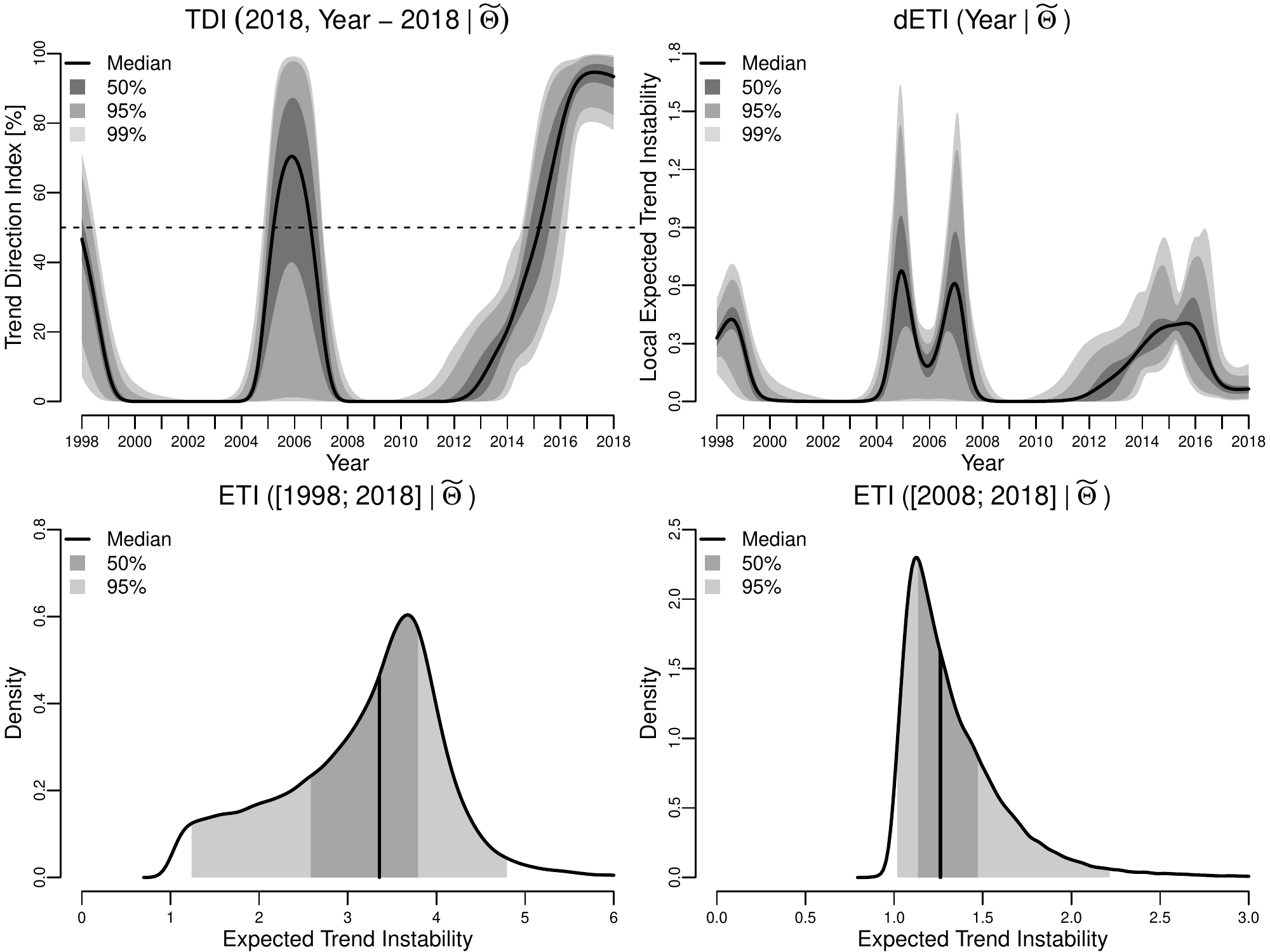}
\caption{Results from fitting the latent Gaussian Process model by Bayesian analysis. The first row shows the posterior distributions of TDI (left) and local ETI (right) with the posterior means in bold and gray areas showing point-wise 95$\%$ and 99$\%$ probability intervals for the posterior distribution. The second row shows densities and probability intervals for the expected trend instability for the 20-year period back-in-time from 2018 (left) and 10 year back-in-time (right).}
\label{fig:bayesFitPlot}
\end{figure}

The results from the two analyses generally agree, but there are two
differences that we wish to address. Both analyses showed a local peak
in trendiness around 2006. In the maximum likelihood analysis this
occurred at 2005.94 with a Trend Direction Index of \(86.47\%\). In the
Bayesian analysis the peak occurs at 2005.87 with a median Trend
Direction Index of \(79.43\%\) and a \(95\%\) posterior probability
interval of \([1.23\%; 97.73\%]\). The added uncertainty estimates
facilitated by the Bayesian estimator shows this trendiness is so
variable that there is no reason to believe in a substantial increase in
proportions at that point in time. This insight could not have been
obtained from the maximum likelihood analysis.

The second difference is that the Bayesian model seems to generally
induce more sluggish estimates due to mixing over the underlying
parameters. This can be seen from the plot of the median local Expected
Trend Instability in Figure \ref{fig:bayesFitPlot} which is generally
lower than its corresponding maximum likelihood point estimates in
Figure \ref{fig:likFitPlot}. This is similarly reflected in the median
ETI estimates in Table \ref{tab:summaries} which are lower than their
values under maximum likelihood. Looking at the posterior distributions
of the covariance parameters \(\theta\) (not shown), we see that this is
mainly a result of not restricting the parameter \(\nu\) to its maximum
likelihood value. The 95\% probability interval of the posterior
distribution of \(\nu\) was \([0.328; 10.743]\) which is highly
right-skewed compared to the maximum likelihood estimate of
\(\widehat{\nu^\text{ML}} = 1.020\).

To understand the effect of \(\nu\) on the smoothness of the fitted
models we can compare the local expected number of crossings by a
Gaussian Process and its derivative at their mean values in the simple
case of a zero-mean process with either the Rational Quadratic or the
Squared Exponential covariance function. In this case the formula in
Proposition \ref{prop:ETIposterior} simplifies immensely, and as shown
in the Supplementary Material the local expected number of
mean-crossings by \(f\) is equal to \(\pi\rho^{-1}\) for both covariance
functions. However, for \(df\) the local expected number of
mean-crossings is equal to \(3^{1/2}\pi^{-1}\rho^{-1}\) for the Squared
Exponential covariance function and
\(3^{1/2}\pi^{-1}\rho^{-1}(1 + \nu^{-1})^{1/2}\) for the Rational
Quadratic covariance function. We note that
\(1 < (1 + v^{-1})^{1/2} < \infty\) for \(0 < \nu < \infty\) and
monotonically decreasing for \(\nu \rightarrow \infty\) with a limit of
one. Therefore, the value of \(\nu\) has no effect on the crossing
intensity of the process itself, but its derivative is always larger
under a Rational Quadratic covariance function compared to the Squared
Exponential covariance function with equality in the limit. A
right-skewed posterior distribution of \(\nu\) therefore favors fewer
crossings of the trend leading to a more stable trend and a smaller
value of the Expected Trend Instability.

\begin{table}[htbp]
\center
\begin{tabular}{l|rrr}
  & Maximum Likelihood & \multicolumn{2}{l}{Bayesian Posterior}\\ \hline
$\text{TDI}(2018, 0)$  & 95.24\% & 93.32\% & $[82.15\%; 98.86\%]$\\
$\text{TDI}(2018, -1)$ & 95.92\% & 94.21\% & $[84.28\%; 99.11\%]$\\
$\text{TDI}(2018, -2)$ & 74.41\% & 77.87\% & $[51.02\%; 94.94\%]$\\
$\text{TDI}(2018, -3)$ & 33.36\% & 44.11\% & $[18.23\%; 69.19\%]$\\
$\text{TDI}(2018, -4)$ & 18.96\% & 20.60\% & $[6.05\%; 31.82\%]$\\
$\text{TDI}(2018, -5)$ & 9.50\% & 6.21\% & $[0.03\%; 22.21\%]$\\ \hline
Crosspoint & 2015.48 & 2015.19 & $[2014.62; 2015.96]$\\ \hline
$\text{ETI}([1998, 2018])$ & 3.68 & 3.36 & $[1.24; 4.79]$\\
$\text{ETI}([2008, 2018])$ & 1.39 & 1.25 & $[1.02; 2.22]$
\end{tabular}
\caption{Summary measures from the Maximum Likelihood and Bayesian analyses. The rows show the estimated Trend Direction Index for 2013 to 2018 and the Expected Trend Instability for the last 10 and 20 years all conditional on data from 1998 to 2018. For the Bayesian analysis posterior medians and 95$\%$ posterior probability intervals are given. Crosspoint is the time during the last $10$ years where $\mathrm{TDI}$ first exceeded $50\%$.}
\label{tab:summaries}
\end{table}

As a final remark to this application we note that the observed data are
proportions and therefore by nature not normally distributed. Commonly
used transformations for proportions towards normality are the isometric
log ratio or the arcsine-square-root or logit functions. In accordance
with Section \ref{sec:gaussianassumptions} we also performed the trend
analysis on the logit transformed outcomes. The results from this
analysis are included in the Supplementary Material and did not give
rise to different interpretations. This is perhaps not surprising as the
observed proportions are far from the boundaries of the parameter space
and consequently the normality approximation is more likely to hold.

\hypertarget{number-of-new-covid-19-positive-cases-in-italy}{%
\subsection*{Number of new COVID-19 positive cases in
Italy}\label{number-of-new-covid-19-positive-cases-in-italy}}
\addcontentsline{toc}{subsection}{Number of new COVID-19 positive cases
in Italy}

As a second application we look at the development of the number of new
COVID-19 positives in Italy since February 24th 2020. Data was updated
each day and made available at the GitHub repository of the Italian
Civil Protection Department (Consiglio dei Ministri - Dipartimento della
Protezione Civile 2020). 90 days had passed when this analysis was
performed.

The data set provides a direct way to assess the COVID-19 disease
progression and to monitor the impact of political initiatives to reduce
disease spread. In this application we wish to assess the statistical
properties of the following questions:

\begin{itemize}
\item[Q1:]{Is the disease spread currently under control or does the number of new positive cases seems to be on the rise?}
\item[Q2:]{How did the Italian government's decision to lockdown most of Italy on March 9th 2020 reflect in the spread of the virus?}
\end{itemize}

For this application we fitted the Gaussian Process model using the
maximum likelihood method and as hyper-parameters we used a constant
mean function and the Rational Quadratic covariance function.

\begin{figure}[htb]
\center\includegraphics{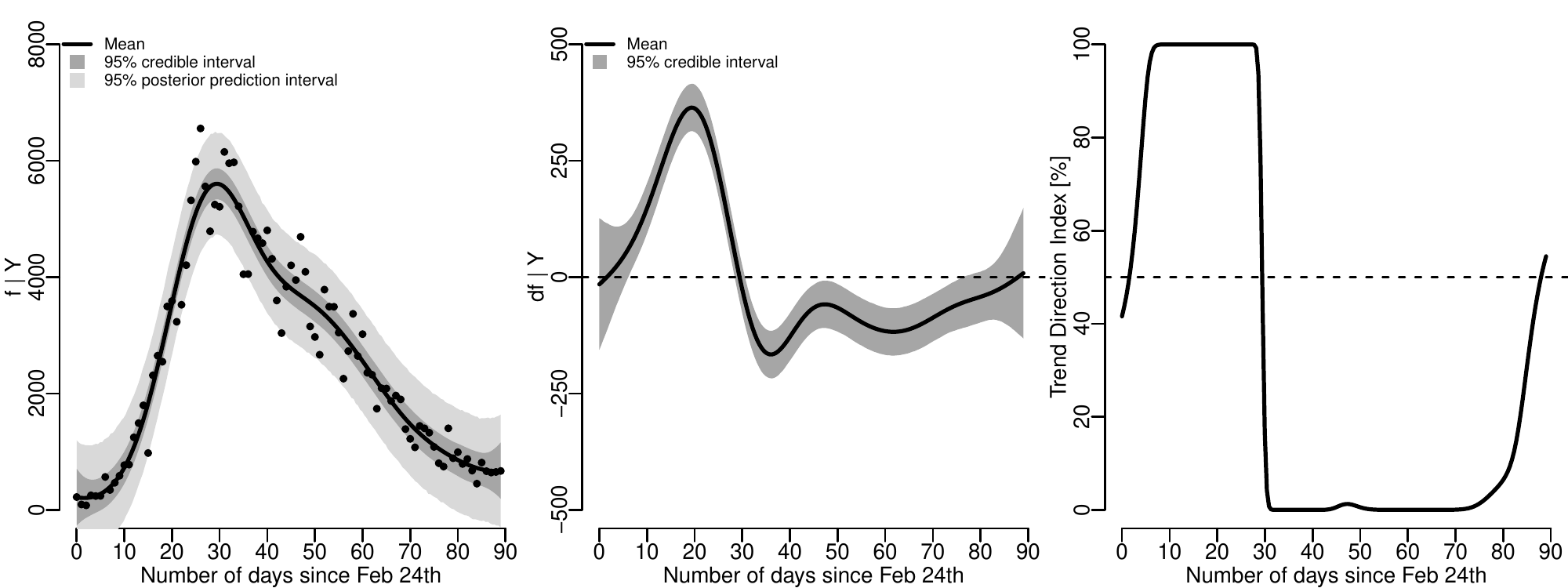}
\caption{Results from the trend analysis of the Italian COVID-19 data. The left panel shows the number of new positives since February 24th along with the posterior mean of $f$ and 95\% credible and posterior prediction intervals. The middle panel shows the posterior distribution of the trend ($df$), and the right panel shows the Trend Direction Index.}
\label{fig:covid19Fit}
\end{figure}

Figure \ref{fig:covid19Fit} shows the result of the analysis. Between
day five and six the Trend Direction Index crossed 95\% probability and
continued to climb to 100\%. After 29 days (March 24th, 2020) the index
proceeded to sharply decrease where it remained for some time. The
Italian government imposed a quarantine on most of Italy from March 9th,
and the sharp drop coincides nicely with the (as of present) expected
incubation period of 2--14 days which roughly reflects the incubation
and testing period for the individuals who contracted the virus before
the March 9th lockdown. After March 24th, the trend was clearly not
positive for a long time.

However, at day 88 (May 22nd, 2020) the Trend Direction Index had again
increased and crossed 50\%, and at the time of this analysis the index
was equal to 54\%. It is therefore currently not certain whether the
number of new positives is increasing or decreasing but a slight
probability favors the former. It should be stressed, that the Trend
Direction Index monitors the \emph{sign} and not the \emph{size} of the
trend and this recent increase of TDI to 54\% may reflect that the
number of new cases is merely starting to level off
(\(df(t) \approx 0\)). The Trend Direction Index can be used as a
monitoring tool to determine if the Italian authorities need to take
extra actions.

\hypertarget{sec:discussion}{%
\section{Discussion}\label{sec:discussion}}

In this article we have proposed two new measures, the Trend Direction
Index and the Expected Trend Instability, in order to quantify the
trendiness of a latent, random function observed in discrete time. Using
a Gaussian Process model for the latent structure we showed how these
indices can be estimated from data in both a maximum likelihood and a
Bayesian framework and provide probabilistic statements about the
monotonicity of the latent development of an observed outcome over time.

Both indices have intuitive interpretations that directly refer to
properties of the latent trend, and our proposed methodology exploits
the assumption of continuity allowing us to calculate a different (and
in many cases more relevant) probability than what can be obtained from
e.g., multiple pairwise comparisons in discrete time.

The comparison with Trend Filtering showed that our proposed method
provides unbiased estimates of the underlying function \(f\) and its
derivative, \(df\) but that we are able to relax the assumptions of
piece-wise linearity of the underlying process, while in addition
providing a probability distribution for the derivative which is used
directly in the Trend Direction Index and which directly provides the
answers to the research questions often posed.

It is worth noting that the indices are scale-free and do not tell
anything about the magnitude of a trend. Consequently, the two indices
should therefore always be accompanied by plots of the posterior of
\(df\). If a prespecified magnitude, \(u\), of a trend is desired, then
the threshold in the definition of the TDI is easily modified to
accommodate this as
\(\mathrm{TDI}_u(t, \delta) = P(df(t + \delta) > u \mid \mathcal{F}_t)\).

In conclusion, we have introduced a method for quantifying the
trendiness of a trend that specifically addresses questions such as
``Has the trend changed?''. Our approach is based on two intuitive
measures that are easily interpreted, provide well-defined measures of
trend behavior, and which can be applied in a large number of
situations. The flexibility of the model is further improved by the
minimum of assumptions necessary to provide about the underlying latent
trend.

\hypertarget{acknowledgements}{%
\section*{Acknowledgements}\label{acknowledgements}}
\addcontentsline{toc}{section}{Acknowledgements}

The authors would like to thank two anonymous reviewers and the
associate editor for providing thoughtful comments that substantially
improved the manuscript.

\hypertarget{bibliography}{%
\section*{Bibliography}\label{bibliography}}
\addcontentsline{toc}{section}{Bibliography}

\hypertarget{refs}{}
\leavevmode\hypertarget{ref-genlasso}{}%
Arnold, Taylor B., and Ryan J. Tibshirani. 2019. \emph{genlasso: Path
Algorithm for Generalized Lasso Problems}.
\url{https://CRAN.R-project.org/package=genlasso}.

\leavevmode\hypertarget{ref-barry1993bayesianchangepoint}{}%
Barry, Daniel, and J. A. Hartigan. 1993. ``A Bayesian Analysis for
Change Point Problems.'' \emph{Journal of the American Statistical
Association} 88 (421): 309--19.

\leavevmode\hypertarget{ref-bassemand1993abrupt}{}%
Basseville, Michele, and Igor V. Nikiforov. 1993. \emph{Detection of
Abrupt Changes: Theory and Application}. Prentice-Hall.

\leavevmode\hypertarget{ref-bergmeir2012use}{}%
Bergmeir, Christoph, and José M Benítez. 2012. ``On the Use of
Cross-Validation for Time Series Predictor Evaluation.''
\emph{Information Sciences} 191: 192--213.

\leavevmode\hypertarget{ref-carlstein1994change}{}%
Carlstein, Edward, Hans-Georg Müller, and David Siegmund, eds. 1994.
\emph{Change-Point Problems}. Vol. 23. Lecture Notes -- Monograph
Series. Institute of Mathematical Statistics.

\leavevmode\hypertarget{ref-carpenter2017stan}{}%
Carpenter, Bob, Andrew Gelman, Matthew D Hoffman, Daniel Lee, Ben
Goodrich, Michael Betancourt, Marcus Brubaker, Jiqiang Guo, Peter Li,
and Allen Riddell. 2017. ``Stan: A Probabilistic Programming Language.''
\emph{Journal of Statistical Software} 76 (1).

\leavevmode\hypertarget{ref-chandler2011statistical}{}%
Chandler, R., and M. Scott. 2011. \emph{Statistical Methods for Trend
Detection and Analysis in the Environmental Sciences}. Statistics in
Practice. Wiley.

\leavevmode\hypertarget{ref-italianCOVID19}{}%
Consiglio dei Ministri - Dipartimento della Protezione Civile,
Presidenza del. 2020. ``COVID-19 Italia - Monitoraggio Situazione.''
2020. \url{https://github.com/pcm-dpc/COVID-19}.

\leavevmode\hypertarget{ref-cramer1967stationary}{}%
Cramer, Harald, and M. R. Leadbetter. 1967. \emph{Stationary and Related
Stochastic Processes -- Sample Function Properties and Their
Applications.} John Wiley \& Sons, Inc.

\leavevmode\hypertarget{ref-esterby1993trenddef}{}%
Esterby, S. R. 1993. ``Trend Analysis Methods for Environmental Data.''
\emph{Environmetrics} 4 (4): 459--81.

\leavevmode\hypertarget{ref-gelman1992inference}{}%
Gelman, Andrew, and Donald B. Rubin. 1992. ``Inference from Iterative
Simulation Using Multiple Sequences.'' \emph{Statistical Science} 7 (4):
457--72.

\leavevmode\hypertarget{ref-gottlieb2012stickiness}{}%
Gottlieb, Andrea, and Hans-Georg Müller. 2012. ``A Stickiness
Coefficient for Longitudinal Data.'' \emph{Computational Statistics \&
Data Analysis} 56 (12): 4000--4010.

\leavevmode\hypertarget{ref-hodrick1997trendfiltering}{}%
Hodrick, Robert J., and Edward C. Prescott. 1997. ``Postwar U.S.
Business Cycles: An Empirical Investigation.'' \emph{Journal of Money,
Credit and Banking} 29 (1): 1--16.

\leavevmode\hypertarget{ref-gptrendStan}{}%
Jensen, Andreas Kryger. 2019. ``GitHub Repository for the Trendiness of
Trends.'' 2019. \url{https://github.com/aejensen/TrendinessOfTrends}.

\leavevmode\hypertarget{ref-kaufman2011efficient}{}%
Kaufman, Cari G, Derek Bingham, Salman Habib, Katrin Heitmann, Joshua A
Frieman, and others. 2011. ``Efficient Emulators of Computer Experiments
Using Compactly Supported Correlation Functions, with an Application to
Cosmology.'' \emph{The Annals of Applied Statistics} 5 (4): 2470--92.

\leavevmode\hypertarget{ref-kim2009trendfiltering}{}%
Kim, Seung-Jean, Kwangmoo Koh, Stephen Boyd, and Dimitry Gorinevsky.
2009. ``\(\ell_1\) Trend Filtering.'' \emph{SIAM Review} 51: 339--60.

\leavevmode\hypertarget{ref-kowal2019dynamicshrinkage}{}%
Kowal, Daniel R., David S. Matteson, and David Ruppert. 2019. ``Dynamic
Shrinkage Processes.'' \emph{Journal of the Royal Statistical Society:
Series B} 81: 781--804.

\leavevmode\hypertarget{ref-mackay1998introduction}{}%
MacKay, D. J. C. 1998. ``Introduction to Gaussian Process.''
\emph{Neural Networks and Machine Learning}.

\leavevmode\hypertarget{ref-micchelli2006universal}{}%
Micchelli, Charles A, Yuesheng Xu, and Haizhang Zhang. 2006. ``Universal
Kernels.'' \emph{Journal of Machine Learning Research} 7 (Dec):
2651--67.

\leavevmode\hypertarget{ref-politiken}{}%
Navne, Helene, Anders Legarth Schmidt, and Lars Igum Rasmussen. 2019.
``Første Gang I 20 år: Flere Danskere Ryger.'' 2019.
\url{https://politiken.dk/forbrugogliv/sundhedogmotion/art6938627/Flere-danskere-ryger}.

\leavevmode\hypertarget{ref-neal2012bayesian}{}%
Neal, R. M. 2012. \emph{Bayesian Learning for Neural Networks}. Lecture
Notes in Statistics. Springer New York.

\leavevmode\hypertarget{ref-quandt1958estimation}{}%
Quandt, Richard E. 1958. ``The Estimation of the Parameters of a Linear
Regression System Obeying Two Separate Regimes.'' \emph{Journal of the
American Statistical Association} 53 (284): 873--80.

\leavevmode\hypertarget{ref-neal1999regression}{}%
Radford, Neal M. 1999. ``Regression and Classification Using Gaussian
Process Priors (with Discussion).'' In \emph{Bayesian Statistics 6:
Proceedings of the Sixth Valencia International Meeting}, edited by A.
P. Dawid José M. Bernardo James O. Berger and Adrian F. M. Smith,
475--501.

\leavevmode\hypertarget{ref-ramdas2016trendfiltering}{}%
Ramdas, Aaditya, and Ryan J. Tibshirani. 2016. ``Fast and Flexible Admm
Algorithms for Trend Filtering.'' \emph{Journal of Computational and
Graphical Statistics} 25 (3): 839--58.

\leavevmode\hypertarget{ref-rasmussen2003gaussian}{}%
Rasmussen, C. E., and C. K. I. Williams. 2006. \emph{Gaussian Processes
in Machine Learning}. MIT Press.

\leavevmode\hypertarget{ref-R-Core-Team:2018aa}{}%
R Core Team. 2018. \emph{R: A Language and Environment for Statistical
Computing}. Vienna, Austria: R Foundation for Statistical Computing.
\url{https://www.R-project.org/}.

\leavevmode\hypertarget{ref-rstan}{}%
Stan Development Team. 2018. ``RStan: The R Interface to Stan.''
\url{http://mc-stan.org/}.

\leavevmode\hypertarget{ref-sst}{}%
The Danish Health Authority. 2019. ``Danskernes Rygevaner 2018.'' 2019.
\url{www.sst.dk/da/udgivelser/2019/danskernes-rygevaner-2018}.

\leavevmode\hypertarget{ref-loo}{}%
Vehtari, Aki, Jonah Gabry, Mans Magnusson, Yuling Yao, and Andrew
Gelman. 2019. ``loo: Efficient Leave-One-Out Cross-Validation and WAIC
for Bayesian Models.'' \url{https://mc-stan.org/loo}.

\leavevmode\hypertarget{ref-williams1998computation}{}%
Williams, Christopher KI. 1998. ``Computation with Infinite Neural
Networks.'' \emph{Neural Computation} 10 (5): 1203--16.

\end{document}